\documentclass[10pt,twocolumn,final,journal]{IEEEtran}

\usepackage{bigstrut,multirow,array,psfig,epsfig,subfigure,amssymb,amsmath,color,url,xr}
\usepackage[ruled,linesnumbered]{algorithm2e}

\begin{document}
\pagenumbering{arabic}


\title{A Multi-Scale Spatial Model for \\
RSS-based Device-Free Localization}

\author{Ossi Kaltiokallio\thanks{Ossi Kaltiokallio is with the Department of Automation and Systems Technology, Aalto University, Helsinki, Finland. E-mail: ossi.kaltiokallio@aalto.fi}, Maurizio Bocca\thanks{Maurizio Bocca and Neal Patwari are with the Department of Electrical and Computer Engineering, University of Utah, Salt Lake City, UT, USA. E-mail: maurizio.bocca@utah.edu, npatwari@ece.utah.edu},  and Neal Patwari \IEEEmembership{Member,~IEEE}}
\maketitle


\begin{abstract}
RSS-based device-free localization (DFL) monitors changes in the received signal strength (RSS) measured by a network of static wireless nodes to locate people without requiring them to carry or wear any electronic device.  Current models assume that the \emph{spatial impact area}, i.e., the area in which a person affects a link's RSS, has constant size.  This paper shows that the spatial impact area varies considerably for each link.  Data from extensive experiments are used to derive a multi-scale spatial weight model that is a function of the \emph{fade level}, i.e., the difference between the predicted and measured RSS, and of the direction of RSS change. In addition, a measurement model is proposed which gives a probability of a person locating inside the derived spatial model for each given RSS measurement. A real-time radio tomographic imaging system is described which uses channel diversity and the presented models. Experiments in an open indoor environment, in a typical one-bedroom apartment and in a through-wall scenario are conducted to determine the accuracy of the system. We demonstrate that the new system is capable of localizing and tracking a person with high accuracy ($\leq0.30$  m) in all the environments, without the need to change the model parameters.
\end{abstract}


\section{Introduction} \label{S:introduction}

\PARstart{I}{n} recent years, wireless sensor networks (WSNs) have been used often for indoor localization. In one of the most common approaches, localization is carried out by measuring the received signal strength (RSS) of the links composing the network. RSS-based localization can be divided into active and passive. Active localization is the practice of locating a person or asset that is carrying an RF tag. Passive localization does not require the person or asset to carry any electronic device, sensor, or tag. The division between these two categories could also be addressed as follows: in the active case, the tracked entity is willing to be located and is co-operating with the system, whereas in the passive case the tracked object is not co-operating with the system, and possibly wants to avoid being located. In this case, the changes in the propagation patterns of the wireless network caused by movements of people can be exploited for localization. These networks are referred to as RF sensor networks \cite{patwari10c}, since the radio frequency (RF) itself is used as the sensing modality. This paper focuses on device-free localization (DFL) \cite{patwari10c}, also referred to as passive localization \cite{youssef07}, sensorless sensing \cite{woyach06}, or RF tomography \cite{patwari08b,kanso09b}.

In recent years, RSS-based DFL has emerged as an attractive technology for passive indoor localization. The technology, which can be used in many applications, such as security and surveillance and assisted living and elder care, has proven its accuracy in open environments \cite{chen11sequential,wilson09a}, in real-world, cluttered environments \cite{youssef07,seifeldin2009nuzzer,kaltiokallio2012b,wilson11fade} and also in through-wall scenarios \cite{wilson10see,zheng2012}. An RSS-based DFL system brings several advantages over other traditional technologies by being able to work in obstructed environments, see through smoke, darkness, and walls, and by avoiding the privacy concerns raised by video cameras. Moreover, a DFL system can be composed of low-cost wireless sensors which measure the RSS, in contrast to e.g. ultra-wideband radars, which are also RF-based and are capable of measuring the amplitudes, phases, and time delays of the multipath signals existing in the radio channel, but are also prohibitively expensive.

Many model-based methods have been already proposed to localize and track people using the temporal variations of RSS.  However, these methods make one or both of the following two assumptions.  First, the movement of a person affects the RSS measurements only when the person is very near the line connecting two communicating transceivers \cite{patwari08b,kanso09b}. Second, when a person's presence is exactly on this line between transceivers, which we call the \emph{link line}, the sensors will strictly observe an attenuation \cite{kanso09b,wilson09a} of the RSS.  In open environments where line-of-sight (LoS) communication among the nodes is dominant and in networks where the distance between the nodes is small, both assumptions are valid. However, for cluttered environments and longer sensor distances, the two assumptions do not apply. In rich multipath environments, the RSS of a link can both increase or remain unchanged as the link line is obstructed. In addition, as the signals propagate via multiple paths from the transmitter to the receiver, it is plausible that a person located far away from the link line affects a subset of multipath components by reflection \cite{liberti96} or scattering \cite{norklit98}, inevitably causing a change in the RSS.  For these reasons, cluttered indoor environments demand more advanced models to characterize the \emph{spatial impact area} in which a person's presence affects the RSS.

In this paper, we demonstrate that the spatial impact area where human-induced RSS changes are measured varies considerably for each link of the RF sensor network. In addition, the spatial impact area is found to depend on the direction of the RSS change, i.e., even for the same link, increases and decreases of the RSS are observed over different spatial areas. We also show that, in addition to a person obstructing the link line, RSS changes vary dramatically depending on the radio frequency channel. Extensive experiments are conducted in two significantly different indoor environments during which more than 26 million RSS measurements are collected. The data are used to build a multi-scale spatial weight model which describes the relationship between the RSS change of a link and the location of the person. The model is built upon the concept of \emph{fade level}, a measure of whether a link is experiencing destructive or constructive multipath interference. In addition, we derive a measurement model which calculates the probability of the person being inside the modeled spatial impact area. Real-time algorithms are developed which exploit the new models and the performance of the system is validated in three different indoor environments, all significantly different from one another. In each environment, the methods proposed in this paper improve considerably the localization accuracy. Furthermore, the more challenging for localization the environment, the greater the enhancement in accuracy provided by the new methods is.

\section{Related Work} \label{S:related_work}
Several works have already shown that human presence and motion alters the way radio signals propagate, enabling the localization and tracking of people. Different measurement modalities, models, algorithms and applications have been proposed, having all as objective to locate people with high accuracy. Most approaches achieve sub-meter localization accuracy. However, a comparison of the results obtained by these systems is difficult since they differ considerably in nodes number, type of indoor environment, hardware, communication protocols, size of the monitored area, just to name a few parameters. In the following, we present the characteristics of some of the already existing DFL systems.

One approach to DFL is to estimate the changes in the RF propagation field of the monitored area and then to image this change field. This image can then be used to infer the locations of people within the deployed network. Estimating the changes in the propagation field is referred to as radio tomographic imaging (RTI), a term coined in \cite{patwari08b}. Several measurement modalities have been proposed for the purpose of RTI. In \cite{wilson09a}, the attenuation of every voxel in the monitored area is estimated using the RSS measurements of many links of the network. Attenuation-based RTI is capable of achieving high accuracy in small and unobstructed deployments, however, in cluttered environments the system loses its capability to locate people.

In \cite{wilson10see}, variance-based RTI (VRTI) is presented, which is capable of localizing and tracking people even through walls of a building. A drawback of VRTI is that it is not capable of localizing stationary people since the measurements are based on a windowed variance of RSS. In addition, VRTI is prone to intrinsic motion which is variance of the RSS that is not caused by people. The problem is confronted in \cite{zhao11noise}, where methods to reduce the noise of VRTI images are presented. Results indicate that the tracking accuracy can be improved significantly by removing the intrinsic noise of the estimated images.

A measurement modality capable of locating both stationary and moving people is presented in \cite{zhao12KRTIa}. Further, the proposed system can achieve high accuracy in open, cluttered, and even in through-wall environments. The system is based on calculating the kernel distance \cite{phillips2011} of two RSS histograms, a long-term histogram representing the RSS measurements when the link line is not obstructed, and a short-term histogram that is capable of capturing the temporal RSS variations when the person is in close proximity to the wireless link.

The works in \cite{kanso09b,kaltiokallio2011,kanso09,kanso09a} also estimate the change in the RF propagation field. However, these works derive the image from a reduced number of links' measurements. In \cite{kaltiokallio2011}, distributed processing is applied to minimize the information transmitted to a central base station, reducing the communication overhead and packet size and increasing energy efficiency. In \cite{kanso09b,kanso09,kanso09a}, the use of compressed sensing, i.e., reconstructing a signal from a reduced number of measurements is investigated. With this method, the estimated images have sharp contrast, while the image formation requires fewer links and can be distributed among the sensors of the network. However, these approaches have been demonstrated only in simple indoor environments and their accuracy in challenging multipath environments has not been proven yet.

The use of channel diversity to enhance DFL accuracy is presented in \cite{kaltiokallio2012a}. The work ranks the different frequency channels used for communication based on two parameters: packet delivery ratio and channel fade level. The results demonstrate that fade level is a more important factor than communication performance when accurate DFL is required. Through channel diversity, the localization accuracy is shown to improve by an order of magnitude compared to a system communicating on a single channel. The multi-channel system in \cite{kaltiokallio2012a} is used in a long-term residential monitoring application in \cite{kaltiokallio2012b}. The work demonstrates the applicability of RSS-based DFL in real domestic environments and the ability of this technology to achieve high localization accuracy over extended periods of time.

A drawback of imaging-based DFL systems is that they first estimate the changed RF-propagation field and then the coordinates of the person. In this two-step process, information can be lost and additional measurement noise can be introduced. Hence, methods to estimate the person's location directly from the RSS measurements are provided in \cite{chen11sequential,wilson11fade,zheng2012}. In \cite{chen11sequential}, a particle filter is applied to simultaneously estimate the location of the sensors and the coordinates of a person moving inside the monitored area. In \cite{wilson11fade}, a fade level skew-Laplace signal strength model and a statistical inversion method are introduced to estimate the location of people. In \cite{zheng2012}, an online learning algorithm is used to determine whether a link is obstructed by a person or not, and a particle filter is used to locate the person. These works rely on sequential Monte Carlo methods to estimate the position of the person. For this reason, due to the computational complexity of particle filters, especially when the number of used particles is high, these systems can not work in real-time.

In the works presented above, the sensors composing the network are deployed around the monitored area. In \cite{zhang2007rf}, the sensors are mounted above the monitored area, i.e. hanging from the ceiling, and a \emph{best-cover algorithm} is applied to estimate the person location when the RSS changes exceed a predefined threshold. The work is extended in \cite{zhang2009dynamic}, where a clustering algorithm is used to track two people.

In this paper, we adapt an imaging based solution \cite{patwari08b,wilson09a,kaltiokallio2012b,wilson10see,zhao11noise,zhao12KRTIa,kaltiokallio2012a} to estimate the changes in the RF propagation field. However, these works assume identical spatial areas for links' area of impact, whereas our algorithm is the first imaging-based DFL approach to set each link's spatial impact area differently. Extensive experiments are conducted to derive a multi-scale spatial weight model which more accurately describes the human induced RSS-changes with respect to the spatial location of people. In addition, a new measurement model is introduced which gives the probability of the person locating inside the modeled area. The multi-scale weight and measurement models are built upon the concept of fade level. Channel diversity is exploited to enhance the accuracy of the system as in \cite{kaltiokallio2012b,kaltiokallio2012a}. However, due to the more accurate weight and measurement models, the system achieves its best accuracy when all the used channels are utilized to estimate the RF propagation field.

\section{Radio Tomographic Imaging} \label{S:rti}

Radio tomographic imaging is the process of estimating the changes in the propagation field of a deployed RF sensor network. Several RTI methods have been already proposed \cite{wilson09a,wilson10see,zhao12KRTIa}. In the following, we describe attenuation-based RTI \cite{wilson09a}, in which the change of the propagation field to be estimated is shadowing, i.e., the attenuation of radio signals.

Let us consider a network that is not occupied by people for a certain period of time. During this time, we can measure the RSS of the $L$ wireless links. The sample mean of link $l$ represents the RSS of the link when this is not affected by people. We denote this as $\bar{r}_l$. We can estimate the attenuation of link $l$ at time $k$ from the change in received signal strength:
\begin{equation}\label{E:shadowing}
    \Delta r_l(k) =  r_l(k) - \bar{r}_l,
\end{equation}
where $r_l(k)$ is the RSS of link $l$ measured at time $k$. Link $l$ experiences shadowing when RSS decreases ($\Delta r_l(k) < 0$). To simplify the notation used throughout the rest of the paper, time $k$ is left out whenever possible. In attenuation-based RTI, the measurement vector $\mathbf{y}$ is the change in links' shadowing, i.e., $\mathbf{y} = [\Delta r_1 \ldots \Delta r_L]^T$. Measurement $y_l$ is a synonym for $\Delta r_l$, however, in Section \ref{S:measurement_based_modeling} we redefine $y_l$.

In RTI, the attenuation of a link is assumed to be a spatial integral of the RF propagation field of the monitored area. Some voxels of the discretized area will affect a particular link's RSS, and some will not. In our discretized model, this corresponds to the fact that each link's change in shadowing is assumed to be a linear combination of the change in voxel attenuations:
\begin{equation} \label{E:voxel_attenuation}
    y_{l} = \sum_{j=1}^{N} a_{lj} x_{j} +  b_{l},
\end{equation}
where $x_{j}$ is the change in attenuation in voxel $j$, $a_{lj}$ the weight of voxel $j$ for link $l$, $b_{l}$ the measurement noise, and $N$ the number of voxels in the image to be estimated.

The weighting $a_{lj}$ represents a spatial model for how voxel $j$'s attenuation affects link $l$'s measurements. In RTI, it is assumed that an elliptical model represents the geometrical relationship between the links and voxels. This model, where the transmitter and receiver are located at the foci, has been effectively applied in \cite{patwari08b,wilson09a,wilson10see,zhao11noise,kaltiokallio2012a}. In this model, the voxels $j$ that are within the ellipse of link $l$ have their weight $a_{lj}$ set to a constant, which is inversely proportional to the square root of the Euclidean distance of the communicating nodes. Otherwise, the weight $a_{lj}$ is set to zero. The weighting is mathematically formulated as follows:
\begin{equation}\label{E:weight}
    a_{lj} =   \begin{cases}
                       \frac{1}{\sqrt{d}} & \text{if } d_{lj}^{tx}+d_{lj}^{rx}<d+\lambda\\
                       0 & \text{otherwise}
                                  \end{cases},
\end{equation}
where $d$ is the distance between the transmitter and receiver, $d_{lj}^{tx}$ and $d_{lj}^{rx}$ are the distances from the center of voxel $j$ to the transmitter and receiver of link $l$, and $\lambda$ is the ellipse excess path length, a parameter that can be used to tune the width of the ellipse.

When all the links of the RF network are considered, the changes in the attenuation field of the monitored area can now be modeled as:
\begin{equation}\label{E:linear_model}
    \mathbf{y} =  \mathbf{A} \mathbf{x} + \mathbf{b},
\end{equation}
where $\mathbf{y}$ and $\mathbf{b}$ are $L \times 1$ vectors representing the measured RSS difference and noise of the $L$ links, $\mathbf{x}$ is the $N \times 1$ radio tomographic image to be estimated, and $\mathbf{A}$ is the weight matrix of size $L \times N$, in which each column represents a single voxel of the image and each row the weight of each voxel for a particular link. The linear model for shadowing loss is based on the correlated shadowing models described in \cite{patwari08b, agrawal09}, and on the work presented in \cite{wilson09a}.

Estimating the image vector $\mathbf{x}$ from the link measurements $\mathbf{y}$ is an ill-posed inverse problem, where the same set of link measurements can lead to multiple different images, i.e., solutions. Therefore, regularization is required. Here, we use a regularized least-squares approach, as in \cite{patwari08b,zhao11noise}:
\begin{equation}\label{E:linear_transformation}
    \hat{\mathbf{x}} =  \mathbf{\Pi}\mathbf{y},
\end{equation}
in which
\begin{equation}\label{E:regularization}
    \mathbf{\Pi} = {(\mathbf{A}^T\mathbf{A}+\mathbf{C}_{x}^{-1}\sigma_{N}^{2})}^{-1}\mathbf{A}^T,
\end{equation}
where $\sigma_{N}^{2}$ is a regularization parameter. The \emph{a priori} covariance matrix $\mathbf{C}_{x}$ is calculated by using an exponential spatial decay:
\begin{equation}\label{E:cov_matrix}
    [\mathbf{C}_{x}]_{ji}=\sigma_{x}^{2}e^{-d_{ji}/\delta_{c}},
\end{equation}
where $d_{ji}$ is the distance from the center of voxel $j$ to the center of voxel $i$, $\sigma_{x}^{2}$ is the variance of voxel attenuation, and $\delta_{c}$ is the voxel's correlation distance. The linear transformation $\mathbf{\Pi}$ can be calculated beforehand, once the positions of the sensors are known, and it has to be computed only once, enabling real-time image reconstruction via (\ref{E:linear_transformation}).

\subsection{Model Errors} \label{S:model_errors}

\begin{figure*}
  \begin{center}
    \mbox{
      \subfigure[]{\includegraphics[width=\columnwidth]{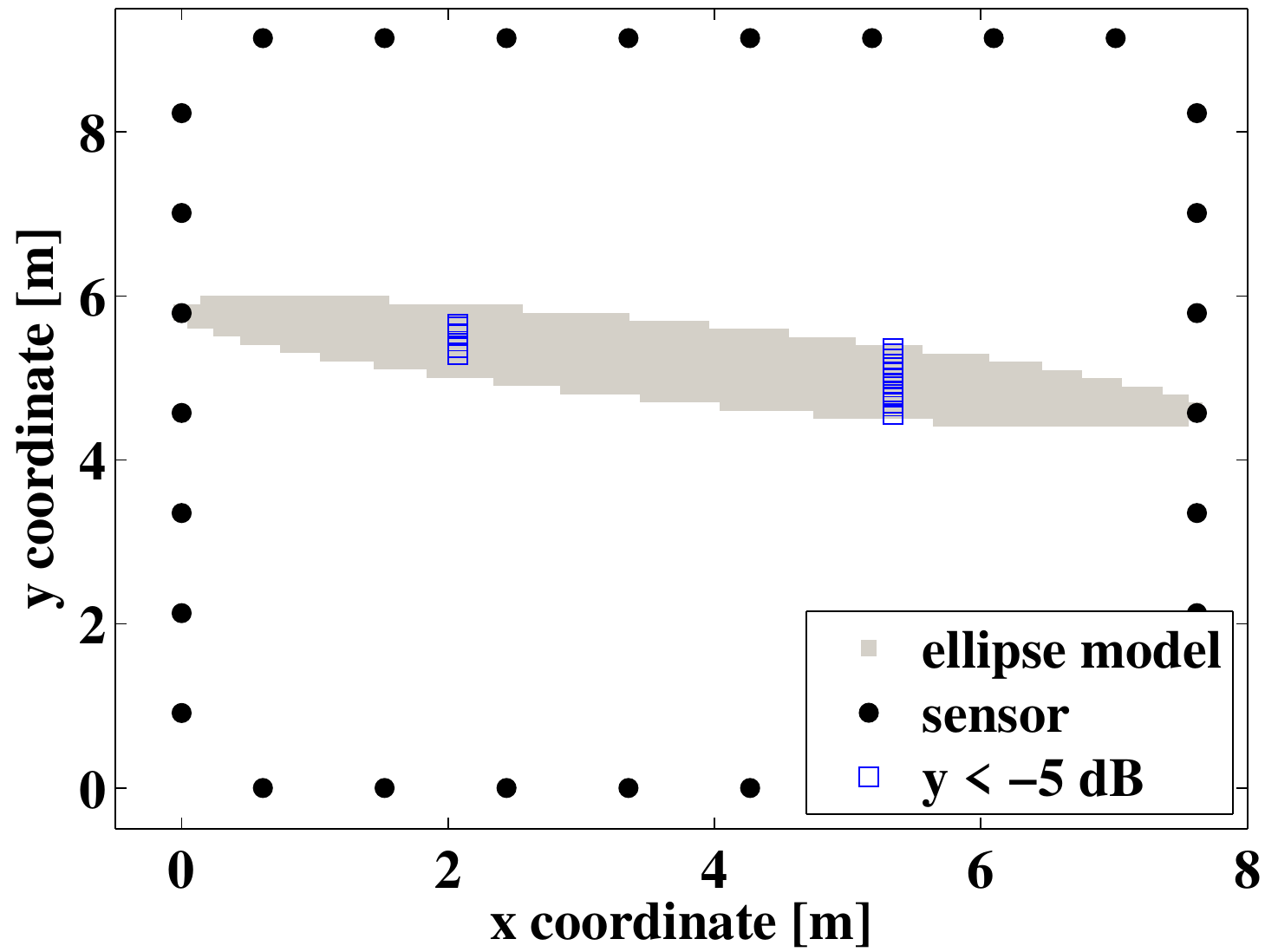}} \quad
      \subfigure[]{\includegraphics[width=\columnwidth]{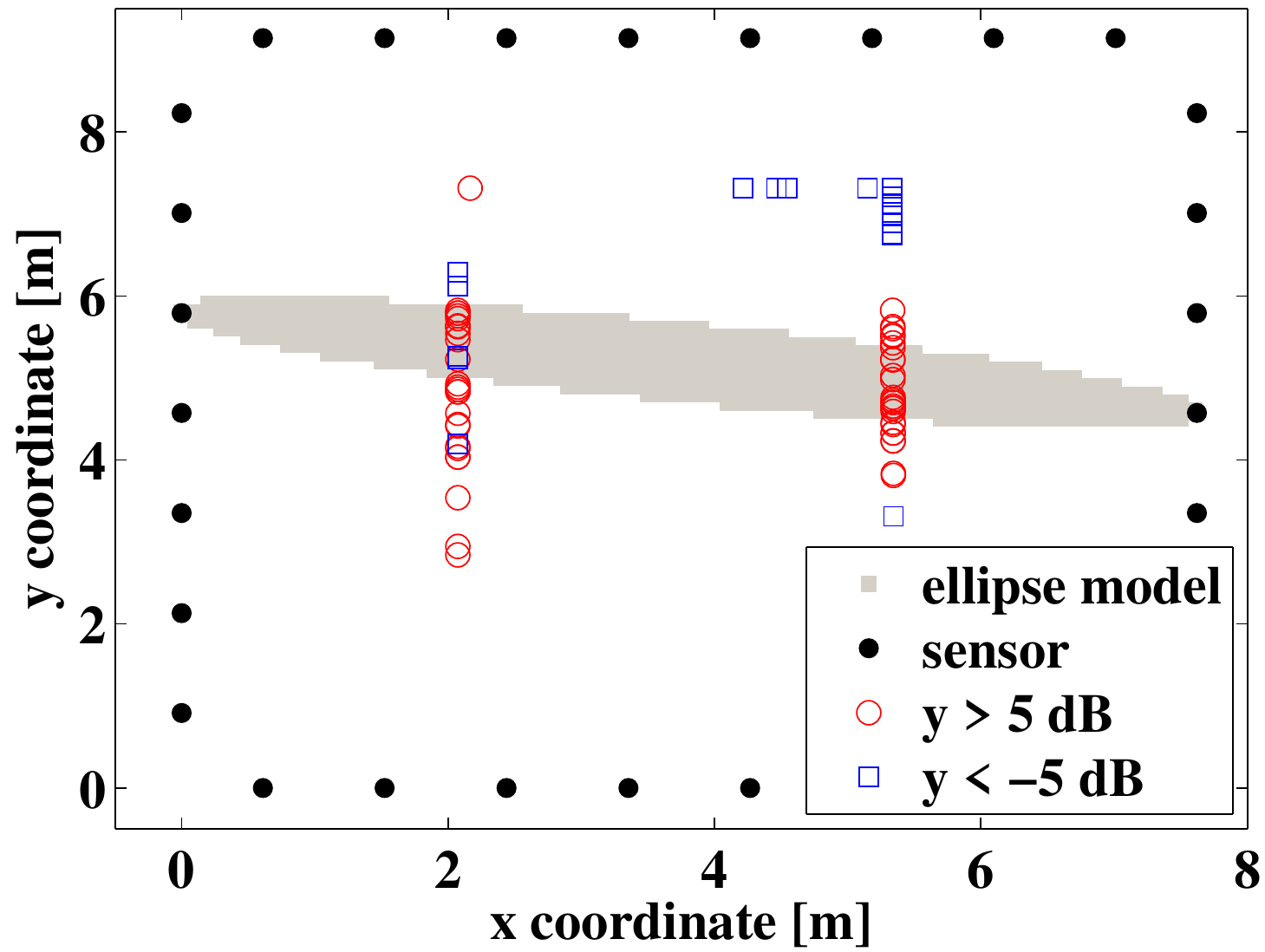}}
      }
    \caption{RSS changes on two different channels for link ([0.0 5.8],[7.6 4.6]) due to human motion. In the figures, the points represent the true location of the person when the link measures a decrease, $y_l < -5$ dB, or an increase, $y_l > 5$ dB, of the RSS. The measurements on channel 15 are shown in (a), whereas the measurements on channel 26 are shown in (b).}
    \label{F:motivation}
  \end{center}
\end{figure*}

The linear model for shadowing loss is based upon two assumptions. First, a person located on the link line causes attenuation. Second, the person affects the RSS measurements only when located near the link line (the value of $\lambda$ is typically set to be small \cite{wilson09a,kaltiokallio2012b,wilson10see}). Let us validate these two assumptions with a test in which 30 nodes communicating on multiple frequency channels are deployed in an unobstructed indoor environment around a square perimeter. During the test, a person walks inside the monitored area along a predefined trajectory.

In Fig. \ref{F:motivation}, the RSS changes of the same link on two different channels are shown. In the figure, the points represent the true position of the person when large RSS changes are measured. For clarity, only large decreases, $y_l < -5$ dB (blue squares), and increases, $y_l > 5$ dB (red circles), are shown. The measurements on channel 15, shown in Fig. \ref{F:motivation} (a), fit well the linear model for shadowing loss. The attenuation of the RSS signal is measured when the person is located near the link line, thus the area where shadowing is observed is accurately predicted by the geometrical ellipse model. However, the measurements of the same link on channel 26, shown in Fig. \ref{F:motivation} (b), indicate that on average an increase in signal strength is measured when the person locates inside the ellipse. In addition, large RSS changes ($\lvert y_l \rvert > 5$ dB) are measured even when the person is located far away from the link line. Consequently, both the measurement and the ellipse weighting models are inaccurate for this link and channel.

\section{Measurement-based Modeling} \label{S:measurement_based_modeling}

In this section we describe the conducted measurement campaign that is used to derive the new multi-scale spatial weight model and the related measurement model. Section \ref{S:link measurements} provides motivation and insight into why more detailed spatial weight models are needed. The section is followed by the presentation of the multi-scale spatial weight model and the new measurement model.

\subsection{Data Collection}\label{S:measurements}

\begin{figure*}
  \begin{center}
    \mbox{
      \subfigure[\quad]{\includegraphics[width=2.2in]{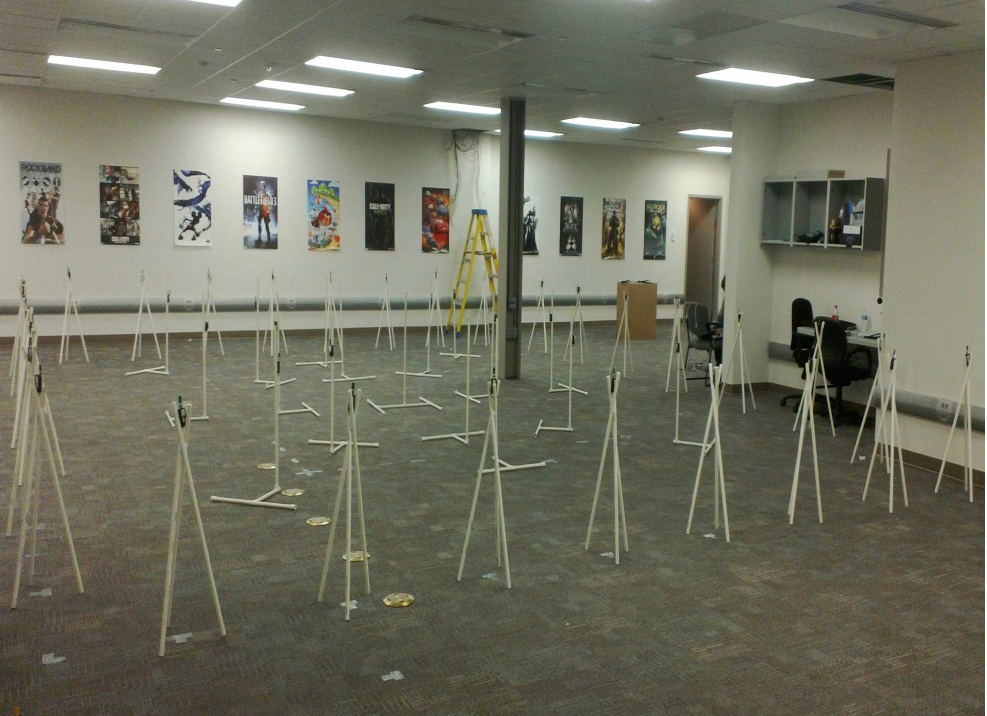}} \quad
      \subfigure[\quad]{\includegraphics[width=2.13in]{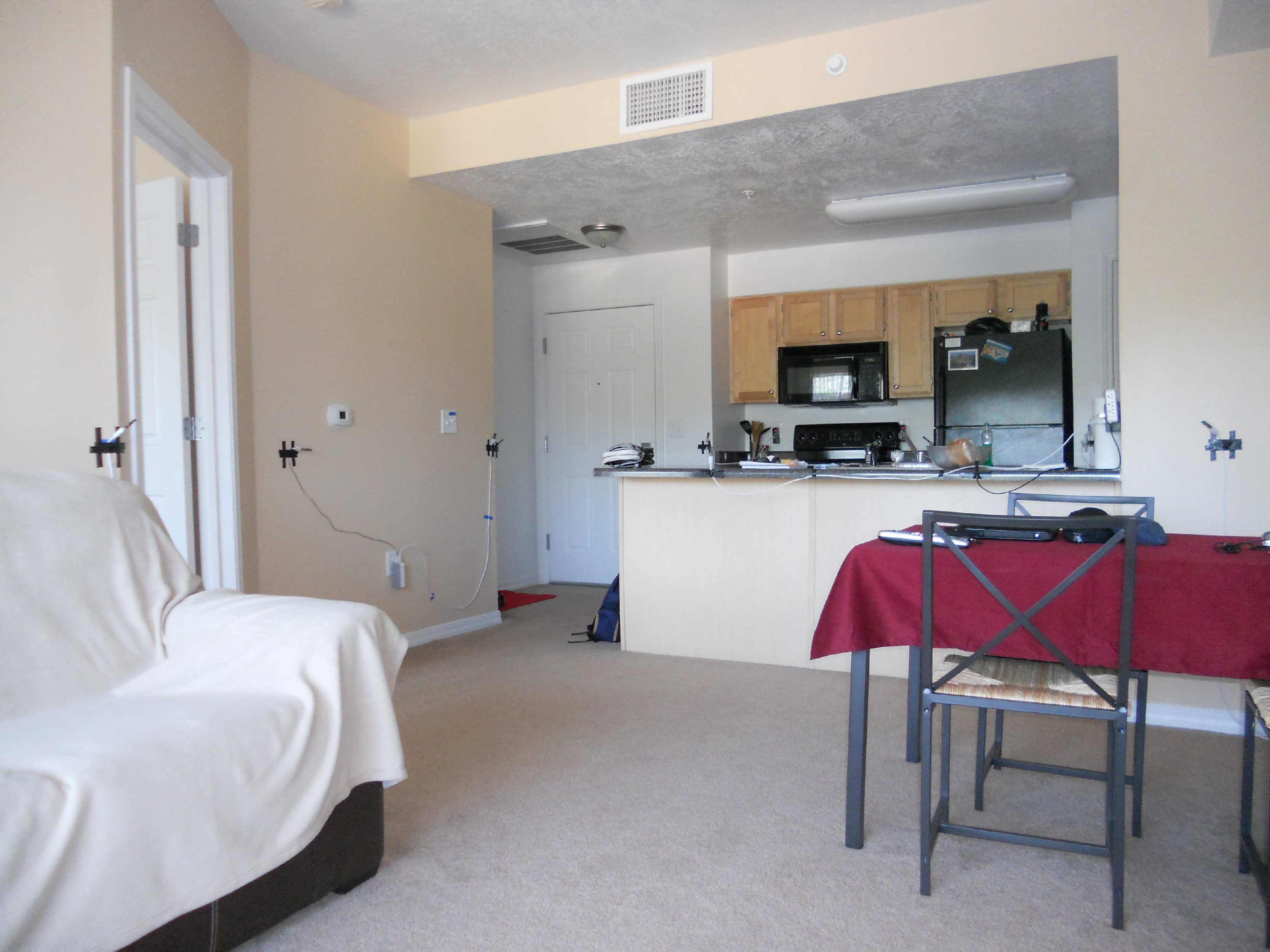}} \quad
      \subfigure[\quad]{\includegraphics[width=2.2in]{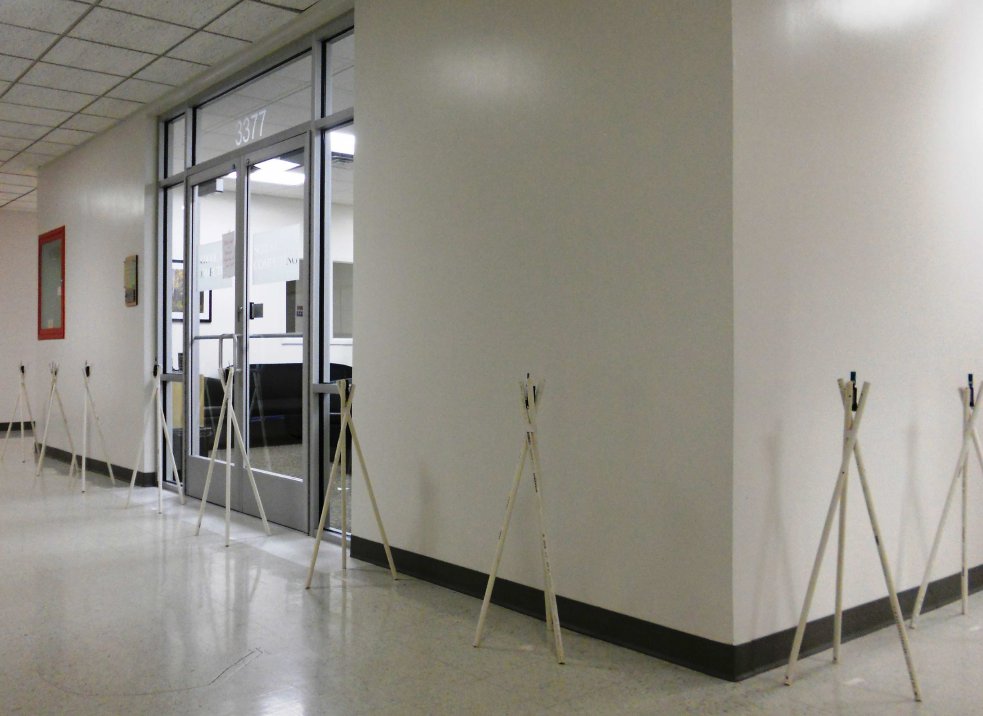}} \quad
      }
    \caption{In (a), an image of the open environment, in (b) the apartment, and in (c) the through-wall scenario is shown.}
    \label{F:environment}
  \end{center}
\end{figure*}

To derive more accurate weight and measurement models, extensive experiments are conducted in two significantly different environments: an open indoor environment (experiment 1), in which all the sensors have LoS communication among each other, and a single-floor, single-bedroom apartment (experiment 2), where multipath propagation is common.

In both experiments, the sensors deployed in the monitored area are Texas Instruments CC2531 USB dongles, set to transmit at the maximum power, i.e. +4.5 dBm \cite{tidongle}. The sensors communicate in TDMA fashion on multiple frequency channels.  The IEEE 802.15.4 standard \cite{802_15_4} specifies 16 channels within the 2.4 GHz ISM band. They are numbered from 11 through 26 and are 5 MHz apart, having a 2 MHz bandwidth. The carrier frequency (in MHz) of channel $c$ is:
\begin{equation}\label{E:normalized_freq}
    f_{c} =  2405+5 \cdot (c-11),\quad c \in [11,26].
\end{equation}

The sequence of transmission is defined by the unique sensor's ID number. In each packet, the sensors include their ID and the most recent RSS measurements of the packets received from the other sensors of the network. If a packet is dropped, the next sensor in the schedule transmits after a back-off time, thus increasing the network's tolerance to packet drops. At the end of each communication cycle, the sensors switch synchronously to the next frequency channel found in a list pre-defined by the user. On average, the time interval between two consecutive transmissions is 2.9 ms. Due to the low latency between transmissions, the human induced changes in RSS of the different channels are correlated. The two experiments are described in detail in the following paragraphs.

In experiment 1, shown in Fig. \ref{F:environment} (a), 30 sensors are deployed on the perimeter of the monitored area (70 m$^2$). The sensors are placed on podiums at a height of one meter. The sensors are programmed to communicate on channels 11, 17, 22, and 26 (no Wi-Fi interference). During the test, a person walks at constant speed along a rectangular path. The trajectory is covered multiple times to collect a sufficient number of RSS measurements. Markers are placed inside the monitored area for the test person to follow, while a metronome is used to set a pre-defined walking pace. In this way, each collected RSS measurement can be associated to the true location of the person.

In experiment 2, shown in Fig. \ref{F:environment} (b), 33 sensors are deployed in a single-floor, single-bedroom apartment (58 m$^2$). Most of the sensors are attached on the walls of the apartment, while a few of them are placed elsewhere, e.g., on the edge of a marble counter in the kitchen or on the side of the refrigerator. The antennas of the sensors are detached from the walls by a few centimeters to enhance the localization accuracy \cite{kaltiokallio2012b}. In the experiment, the sensors are programmed to communicate on channels 15, 20, 25, and 26 in order to avoid the interference generated by several coexisting 802.11 (Wi-Fi) networks, which would increase the floor noise level of the RSS measurements \cite{Srinivasan2006}. As in experiment 1, a person walks along a pre-defined path at constant speed several times.

In both environments, the tests are repeated three times: two of the tests are used to build the new models presented in this work, whereas one of the tests is used to validate the improved accuracy obtained by using the new models. To derive the new models, more than 26 million RSS measurements are collected. Each measurement is associated to the true location of the person. Since the new models are based on the changes in RSS, an initial calibration at the beginning of each test is performed while the monitored area is not occupied by people.

\subsection{Link Measurements}\label{S:link measurements}

\begin{figure}
  \begin{center}
    \includegraphics[width=\columnwidth]{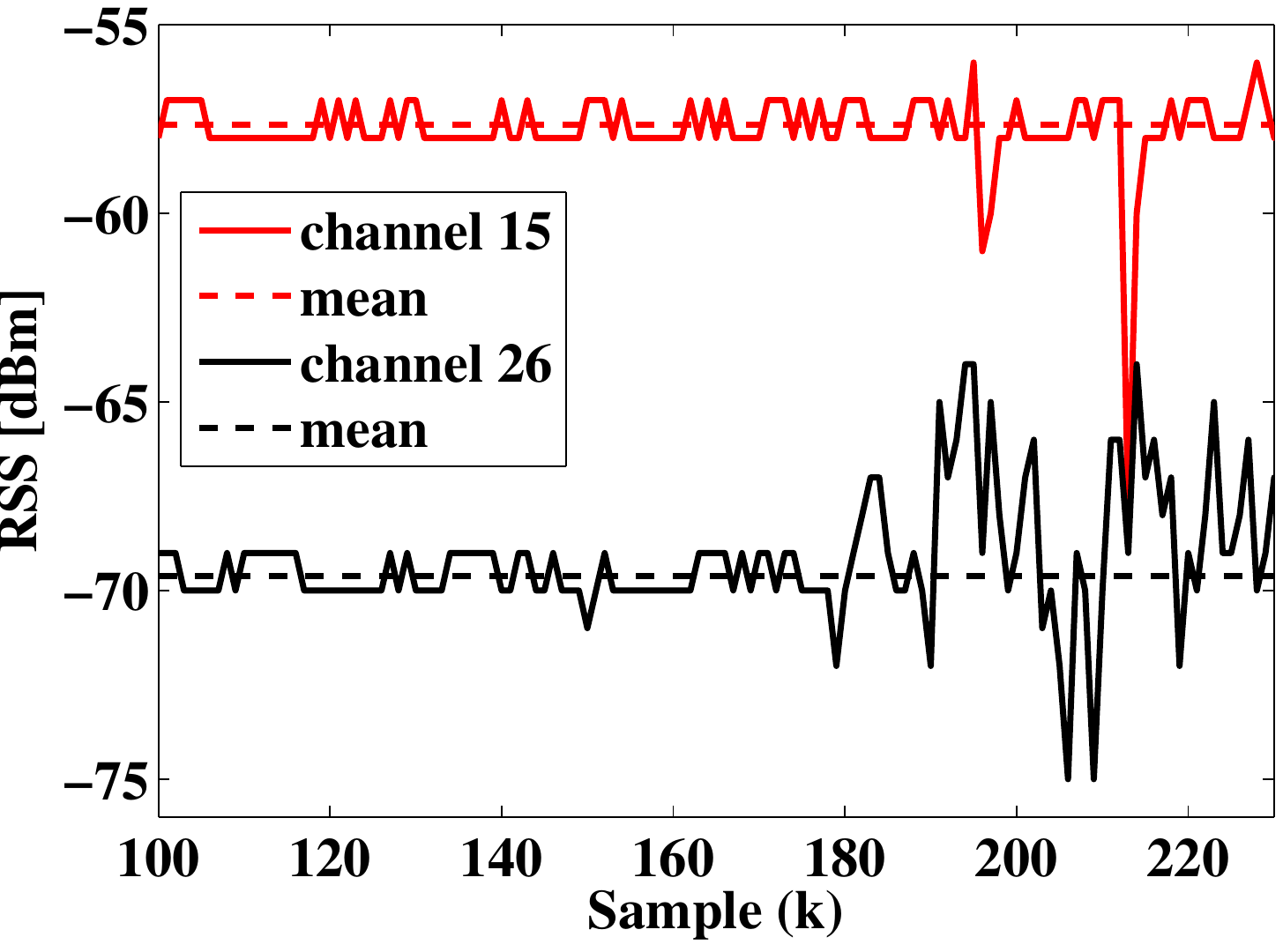}
    \caption{The RSS measurements of a single link on two channels in different fade level conditions.}
    \label{F:link_measurements}
  \end{center}
\end{figure}

People found in close proximity of a wireless link cause changes in the RSS by shadowing, reflecting, diffracting, or scattering a subset of its multipath components \cite{patwari10c,wilson10see}. In open environments in which LoS communication among the sensors is dominant, the sensors measure on average a decrease in RSS when a person is located on the link line. The attenuation of the RSS signal is caused by shadowing. This phenomenon has been successfully applied in attenuation-based DFL \cite{patwari08b,kanso09b,chen11sequential,wilson09a,kaltiokallio2011,kaltiokallio2012a}.

In cluttered environments in which multipath is common, the changes in RSS due to the presence of people become more unpredictable. As the link line is obstructed, the RSS can increase, decrease, or remain unchanged \cite{wilson11fade,wilson10see,kaltiokallio2012a}. In addition, a person can impact the RSS even when located far away from the line connecting the sensors \cite{kaltiokallio2012a}. Temporal variations of the RSS have been successfully applied to track moving people \cite{wilson10see} and also to locate and track both stationary and moving people \cite{wilson11fade} in multipath rich environments.

The statistics of steady-state, narrow-band fading are related to the changes in RSS due to the human presence as described in \cite{wilson11fade}. In this work, the authors define the relation using fade level, a continuum between two extremes: \emph{anti-fade} and \emph{deep fade}. A link in a deep fade experiences destructive multipath interference and will measure on average an increase in RSS when obstructed. In addition, links in deep fade are also sensitive to movement far from the link line, measuring high variance of the RSS. On the contrary, links in anti-fade experience constructive multipath interference. On average, these links remain unchanged when people are moving far from the link line and measure a decrease in RSS when the link line is obstructed. Due to these characteristics, it is the links in anti-fade that are the most informative for RSS-based DFL, since for them the area where the person affects the RSS is small, i.e., a narrow ellipse having TX and RX at the foci. Furthermore, for anti-fade links the direction of RSS change is predictable, thus easier to model.

\begin{figure*}
  \begin{center}
    \mbox{
      \subfigure[\quad]{\includegraphics[width=\columnwidth]{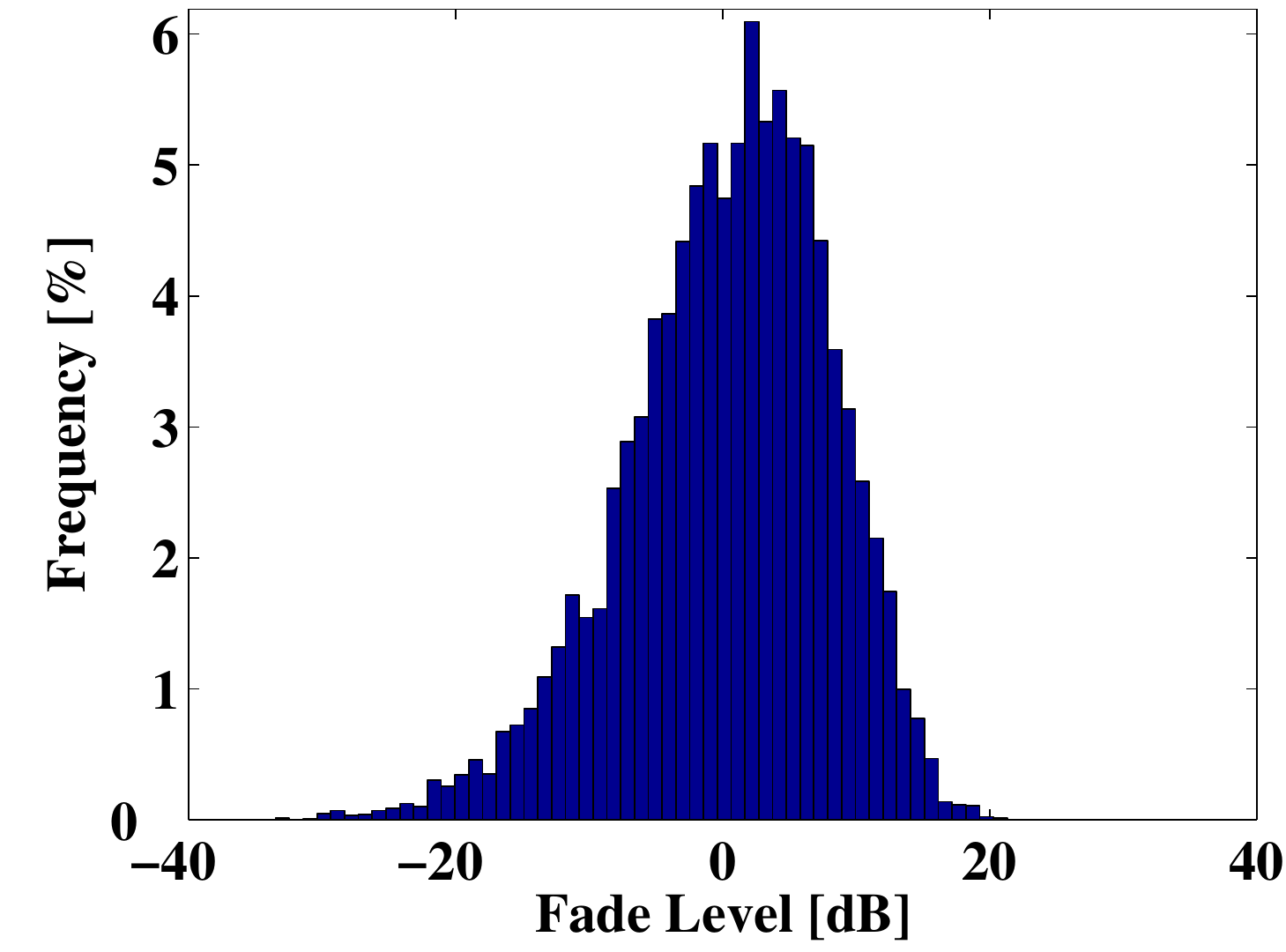}} \quad
      \subfigure[\quad]{\includegraphics[width=\columnwidth]{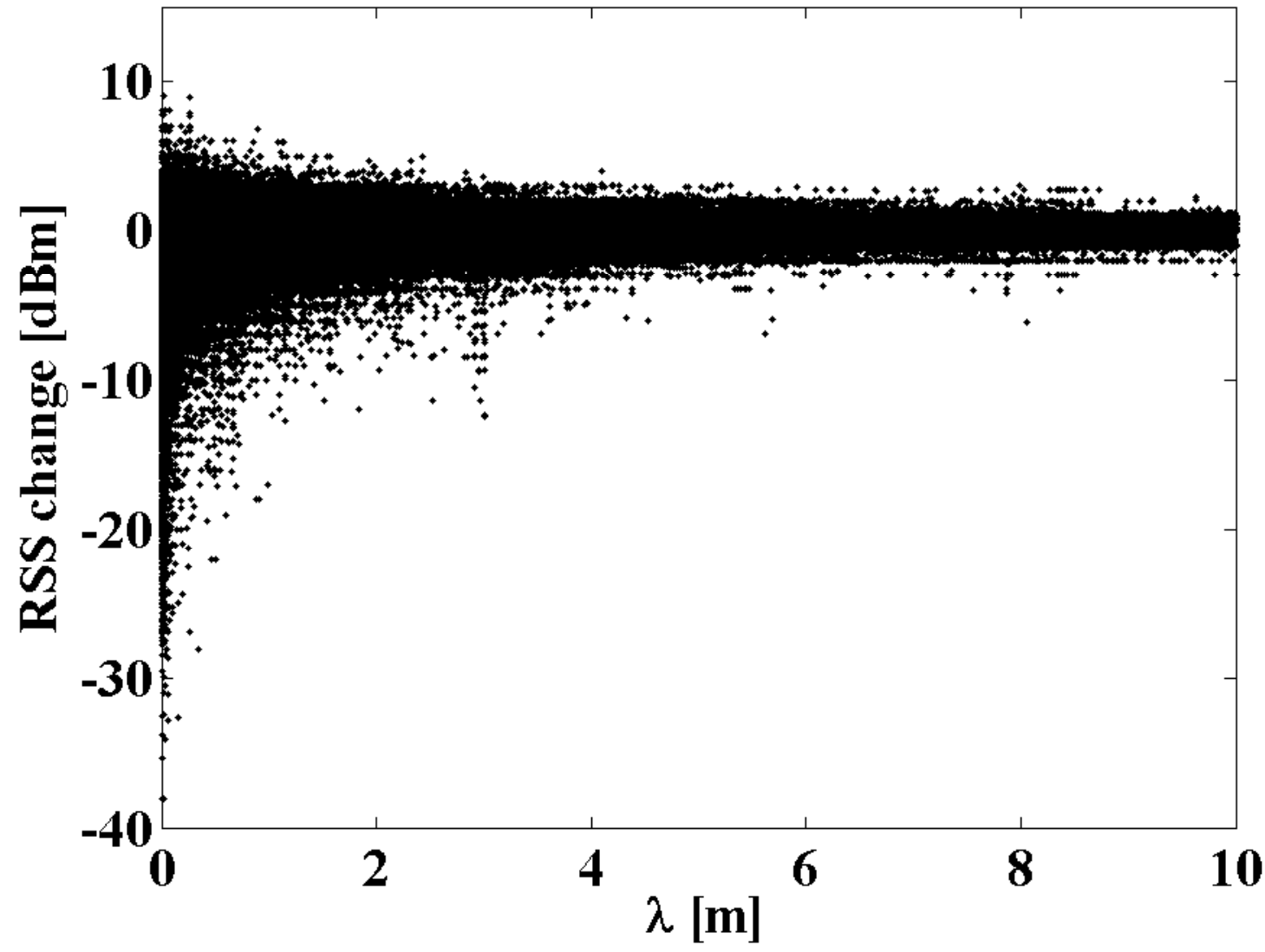}} \quad
      }
    \caption{In (a), the combined distribution of the fade levels in experiments 1 and 2. RSS measurements of links that have a fade level of $F \in [7.86, 9.42]$ dB as a function of $\lambda$ are shown in (b).}
    \label{F:fade_level_measurements}
  \end{center}
\end{figure*}

To illustrate the effect of fade level on the temporal RSS changes, the measurements of a single link on two different channels are plotted in Fig. \ref{F:link_measurements}. The link and channels considered in Fig. \ref{F:link_measurements} are the same as the ones shown in Fig. \ref{F:motivation}. The monitored area is empty in the beginning of the experiment until a person enters it ($k$ = 185) and starts walking along a predefined trajectory. The link line is intersected two times ($k$ = 196 and $k$ = 213) within the considered period of time. The difference in fade level of the two channels is almost 12 dB. The channel in anti-fade (channel 15) measures attenuation when the link line is crossed. On the contrary, the deep fade channel (channel 26) measures an increase in signal strength as the link line is crossed. In addition, the measurements of the deep fade channel fluctuate much more than the measurements of the anti-fade channel when the person is moving off the link line.

The measurements in Figs. \ref{F:motivation} and \ref{F:link_measurements} and previous works found in the literature \cite{wilson11fade,wilson10see,kaltiokallio2012a} show that the RSS measurements are very different for links in different fade levels. This motivates us to further explore how the fade level of the links relates to the spatial impact area of the person, the measured RSS and its direction of change. In this paper, we propose more sophisticated models, derived from experimental data, which describe more accurately the area where human motion induces RSS changes and the temporal variations of the RSS.

\subsection{Fade Level}\label{S:fade_level}

In obstructed environments, where non-LoS communication is dominant, the signals travel via multiple paths from transmitter to receiver, a phenomenon called multipath propagation. At the receiver, a phasor sum of the radio waves impinging on the antenna determine the strength of the received signal. This phasor sum may be constructive or destructive, depending on the phase of the waves. This results in a RSS that is a function of the center frequency and position in space, an effect called multipath fading, which causes a significant deviation from the theoretical radio propagation model. The difference between the radio propagation model and the mean RSS of link $l$ on channel $c$, $\bar{r}_{c,l}$, is what we call fade level, and can be estimated as follows:
\begin{equation} \label{E:fade_level}
    F_{c,l} = \bar{r}_{c,l}-P(d,c),
\end{equation}
where $P(d,c)$ is a model for the RSS vs. distance and channel.

In a wireless network, the signal strength measured at the receiver can be modeled with the log-distance path loss model \cite{rappaport96}:
\begin{equation} \label{E:path_loss}
    P(d,c) = P_0(c)-10\eta\log_{10}\frac{d}{d_0},
\end{equation}
where $P_0(c)$ is what we call transmit power-normalized reference loss on channel $c$ and $\eta$ is the path loss exponent. The transmit power-normalized reference loss is defined as:
\begin{equation} \label{E:TX_normalized}
    P_0(c)=P_T(c)-P_0,
\end{equation}
where $P_T(c)$ is the normalized transmit power on channel $c$ and $P_0$ the reference loss at a short reference distance $d_0$. The normalized transmit power $P_T(c)$ is a linear function of the used frequency channel, which must be derived from experimental data. For the wireless sensor used in the experiments \cite{tidongle}, the lower frequency channels measure lower RSS values than the higher ones, because of the differences in antenna impedance matching across a wide frequency band \cite{tidongleantenna}. We find the linear relationship $P_T(c)=(0.1452 \cdot c+1.7332)$ dB matched the measured transmit power closely. Normalization is required to avoid bias in the fade level estimates.

To derive the path loss exponent $\eta$ in (\ref{E:path_loss}), a calibration is performed at the beginning of each test. At the end of this period $\bar{r}_{c,l}$ for every link and channel is calculated. Since the distance between the nodes is known, a least-squares linear fit can be used to determine the path loss exponent $\eta$. In an indoor environment, it is typical that some sensors measure on average RSS lower than what predicted by (\ref{E:path_loss}), whereas other sensors have a positive fade level. The combined distribution of the fade levels in experiments 1 and 2 is shown in Fig. \ref{F:fade_level_measurements}(a).

\subsection{Multi-Scale Spatial Weight Model}\label{S:ellipse_model}

\begin{figure*}[t]
  \begin{center}
    \mbox{
      \subfigure[]{\includegraphics[width=\columnwidth]{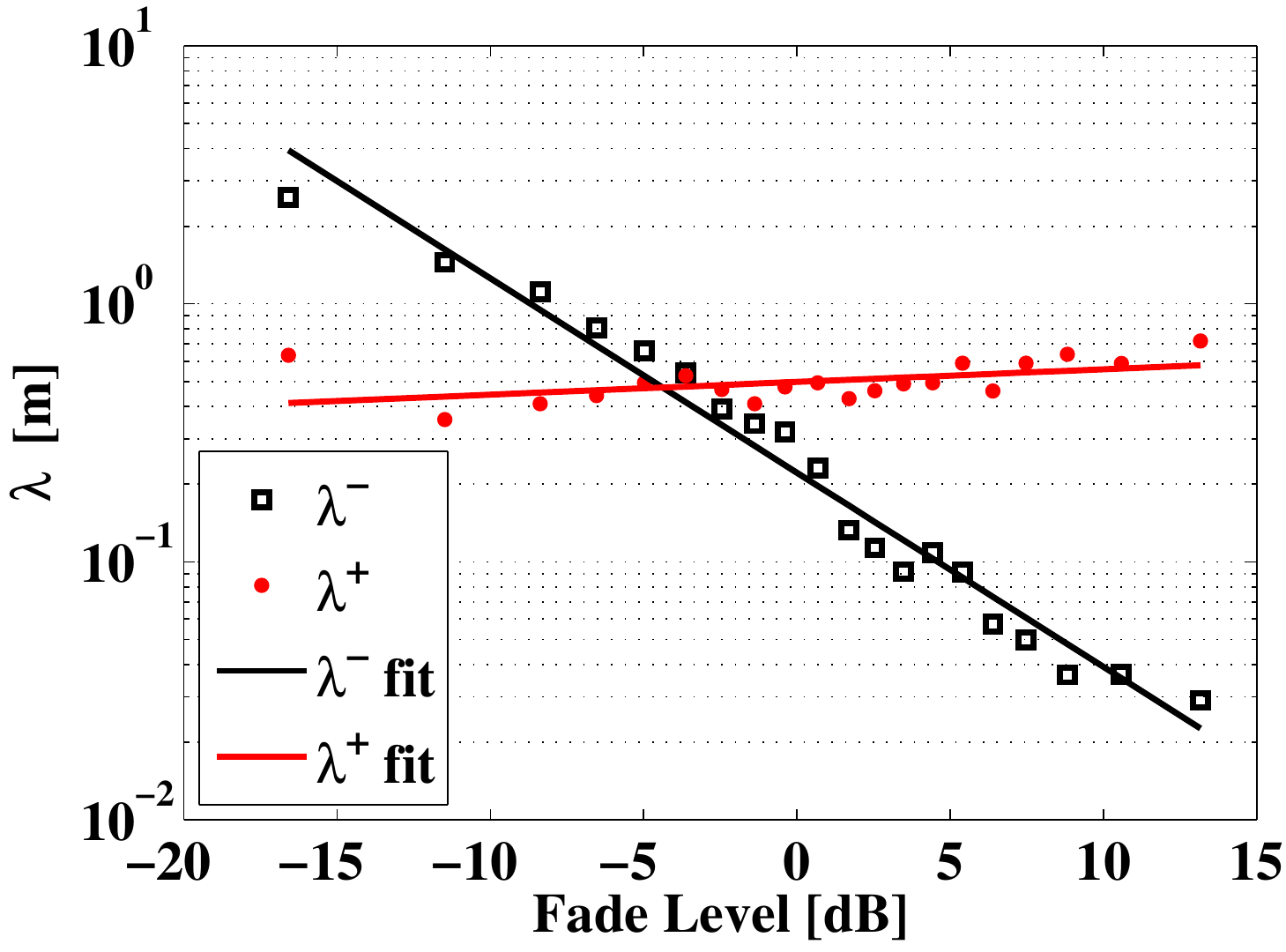}} \quad
      \subfigure[]{\includegraphics[width=\columnwidth]{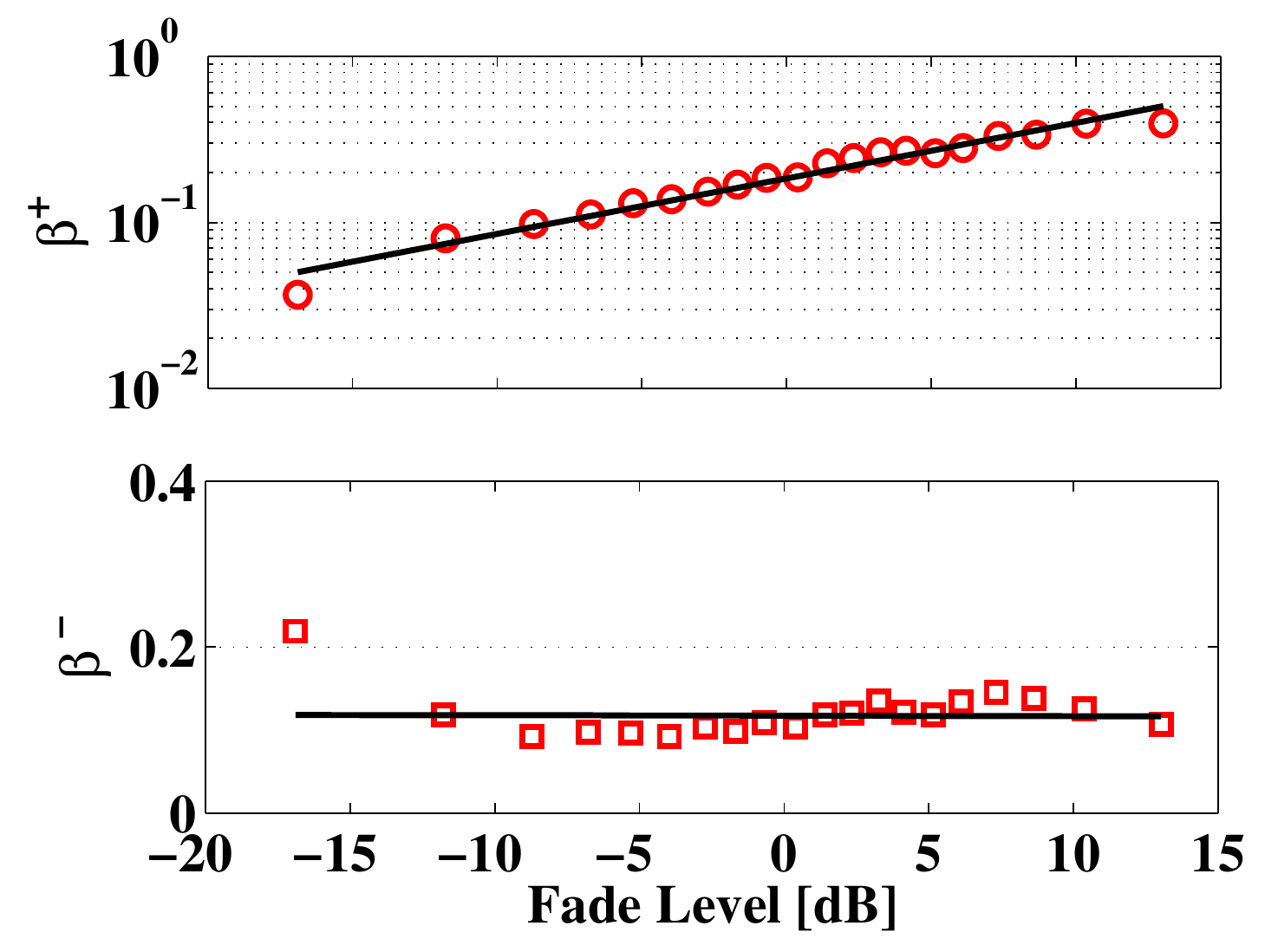}}
      }    
    \caption{In (a), measured and modeled $\lambda^-$ and $\lambda^+$ values as a function of the fade level. In (b), the measured and modeled $\beta^-$ and $\beta^+$ values shown as a function of fade level.}
    \label{F:model_fits}
  \end{center}
\end{figure*}

During the measurement campaign, we determine two measures while the person is moving inside the monitored area. First, the change in RSS for each time instant $k$, $\Delta r_{c,l}(k)$. Second, the minimum excess path length for each link at each time instant $k$, i.e., $\lambda_l(k) = d_{lj}^{tx} + d_{lj}^{rx} - d_l$, so that the person's location is on the perimeter of the ellipse. The excess path length $\lambda_l$ measures the distance of the person from the link line.

We divide the RSS measurements of the experiments in $20$ equally likely bins each containing the same amount of measurements based on their fade level. Fig. \ref{F:fade_level_measurements}(b) shows the change in RSS as a function of $\lambda$ for the links of bin $F \in [7.86, 9.42]$ dB. For each bin we determine the 5th-percentile of measurements that account for the largest decreases in RSS. For this set of measurements, we derive the median excess path length, which we denote as $\lambda^-$. Correspondingly, we calculate $\lambda^+$ for the 5th-percentile of measurements that account for the largest increases in RSS. The values of $\lambda^-$ and $\lambda^+$ for the different fade level bins are shown in Fig. \ref{F:model_fits} (a).

The results indicate that links in a deep fade, i.e. with a negative fade level, measure decreases in RSS within a large area. The excess path length of the ellipse decreases as the fade level increases, confirming the fact that, when RSS attenuation is observed, anti-fade links are more informative for DFL. The opposite is true when an increase in RSS is measured. In this case, links in a deep fade measure the human induced RSS change within a smaller area than the anti-fade links. However, the difference in $\lambda^+$ for deep fade and anti-fade links is relatively small compared to the difference in $\lambda^-$ values.

An exponential decay model accurately fits the $\lambda^-$ values. Correspondingly, an exponential growth model is used to fit the $\lambda^+$ values. For both, a least-squares fit is used to derive the model parameters.  Now, $\lambda$ can be determined separately for each link, frequency channel and direction of RSS change based on the fade level of the link:
\begin{equation}\label{E:lambda_model}
    \lambda_{c,l}^{\delta} = b_{\lambda}^{\delta} \cdot e^{F_{c,l}/{k_{\lambda}^{\delta}}},
\end{equation}
where $\delta$ indicates the direction of RSS change, i.e., $\delta$ is $-$ for measured decreases in signal strength and $+$ for the measured increases. Table \ref{T:ellipse_model_parameters} shows the values of the parameters of the ellipse model.

\begin{table}
    \caption{Ellipse model parameters} 
        \centering 
        \begin{tabular}{c c c} 
        \hline\hline\ 
        Parameter & $\delta$ is $-$ & $\delta$ is $+$ \\
        \hline  
        $k_{\lambda}^{\delta}$ & -5.7874 & 102.7284 \\ 
        $b_{\lambda}^{\delta}$ & 0.2112 & 0.5016 \\ 
        \hline 
        \end{tabular}
        \label{T:ellipse_model_parameters} 
\end{table}

\begin{figure*}[htbp]
  \begin{center}
    \mbox{
      \subfigure[\quad]{\includegraphics[width=\columnwidth]{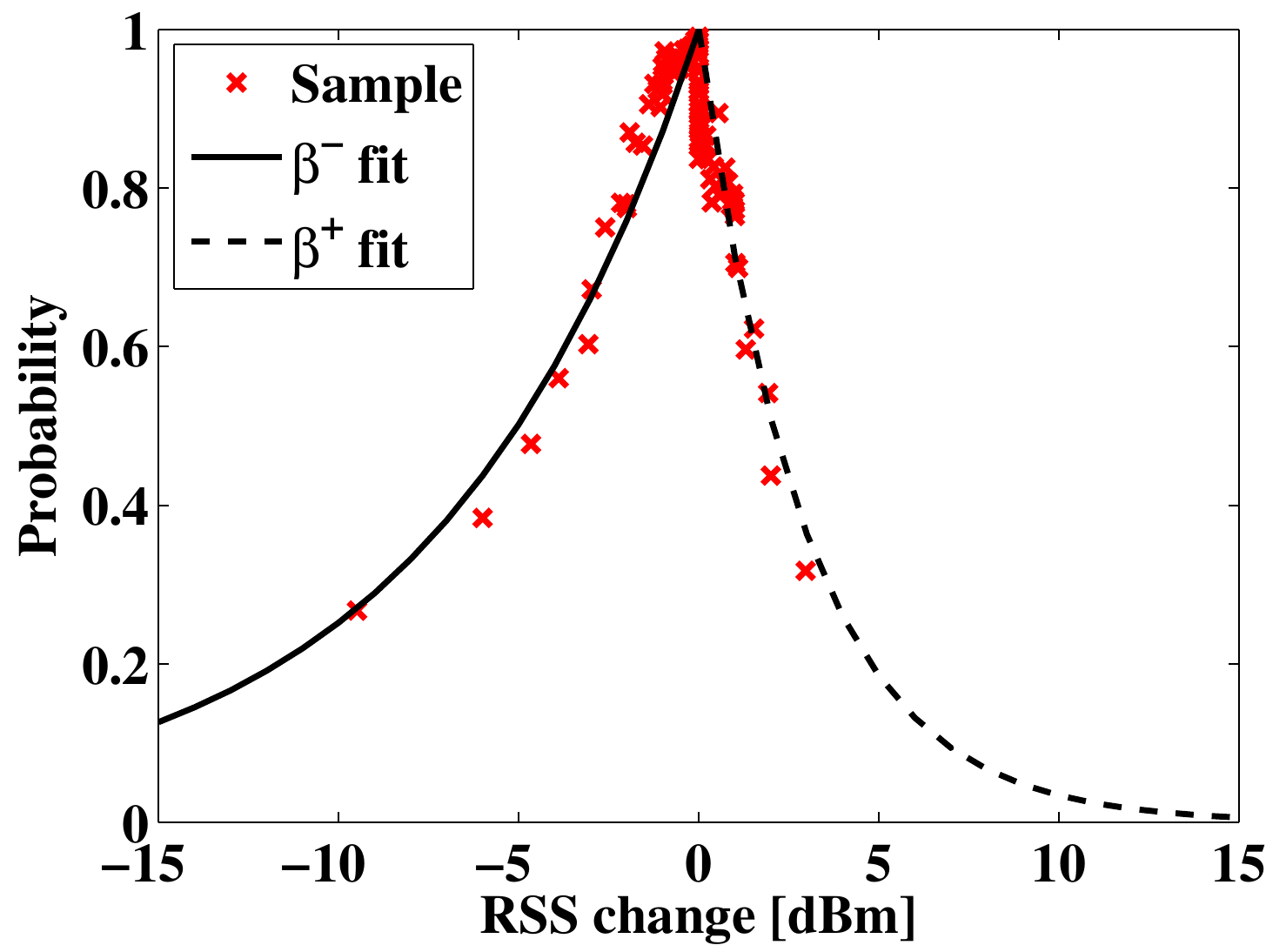}} \quad
      \subfigure[\quad]{\includegraphics[width=\columnwidth]{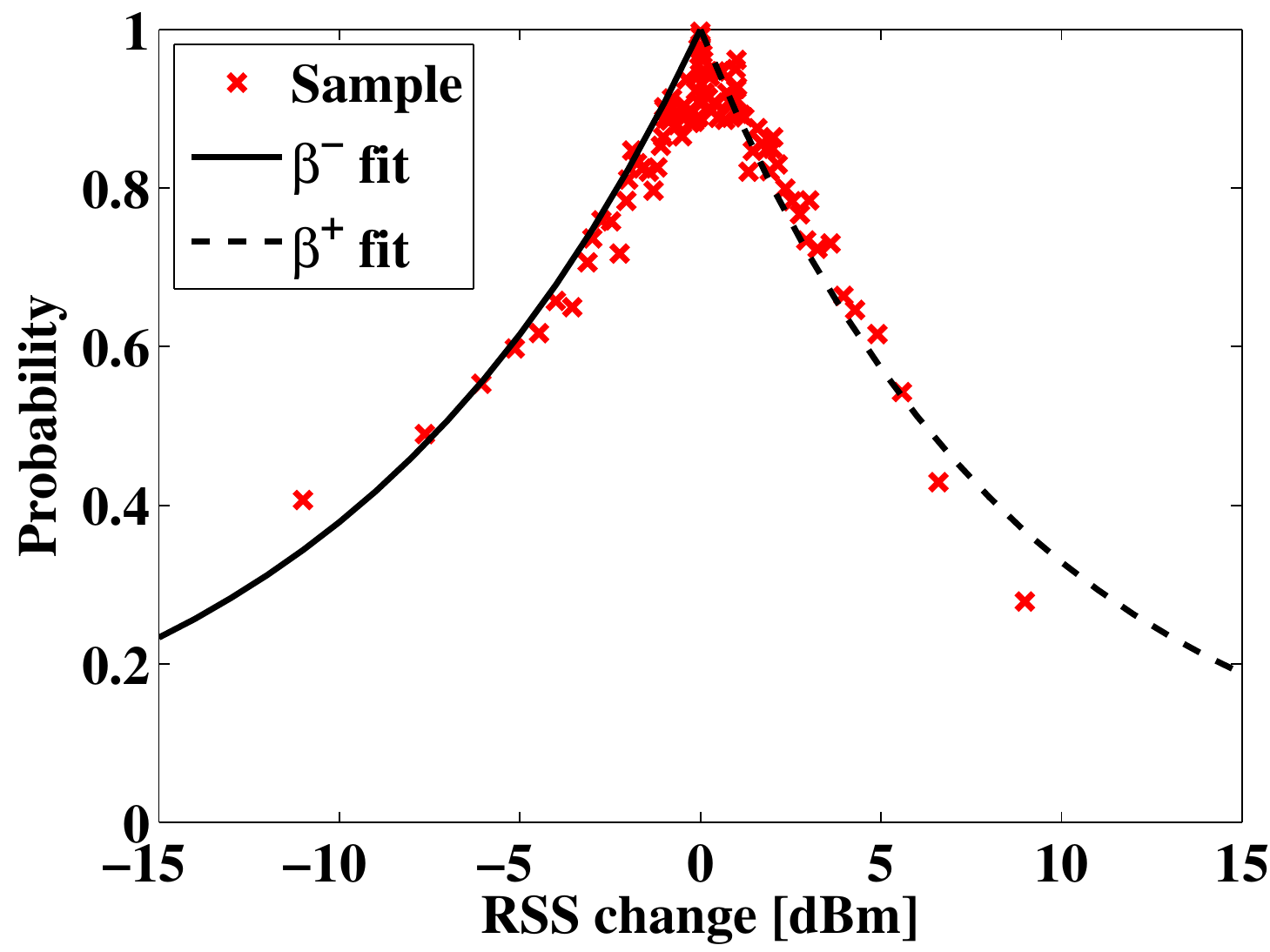}} \quad
      }
    \caption{The probability of the person locating outside the modeled ellipses as a function of measured RSS change shown in (a) and (b). In (a), the fade level bin $F \in [7.86, 9.42]$ dB is shown and in (b), bin $F \in [-13.70, -10.13]$ dB is illustrated.}
    \label{F:pr_fit}
  \end{center}
\end{figure*}

\subsection{Measurement Model}\label{S:measurement_model}

In the previous section it was shown that the area in which a person affects the link's RSS measurements strongly depends on the fade level of the link and the direction of the RSS change. In this section, we derive a measurement model to determine the probability of the person being within the area predicted by the new multi-scale weight model. The new measurement model is based on both the magnitude and direction of RSS change, and fade level of the link.

The RSS measurements are organized as in the previous section. This time, the measurements of each fade level bin are further divided into $100$ bins so that each bin contains $\Delta r_{c,l}$ measurements having approximately the same value. For each bin, we estimate the probability of the person being outside the ellipse. Figure \ref{F:pr_fit}(a) shows the probabilities of the bin $F_l \in [7.86, 9.42]$ dB (i.e., anti-fade links) that the person is located outside the modeled ellipses for each $\Delta r_{c,l}$. The negative x-axis of the figure shows the probability of the person being outside the ellipse $\lambda^{-}$ when a decrease in signal strength is measured. Correspondingly, the positive x-axis shows the probability of the person being outside the ellipse $\lambda^+$ when an increase in RSS is measured. The probabilities of bin $F_l \in [-13.70, -10.13]$ dB (i.e., deep fade links) are illustrated in Fig. \ref{F:pr_fit}(b) for comparison.

The probability that the person is located inside the modeled ellipse depends on the magnitude of $\Delta r_{c,l}$. The larger the RSS change is, the more likely it is that the person is located inside the modeled ellipse. It is again the anti-fade links that are more informative and trustworthy. For example, when a link measures a $10$ dBm attenuation, the anti-fade link ($F_l = 8$ dB) has a 69$\%$ probability that the person is inside the modeled ellipse. In comparison, the probability is the same with the deep fade link ($F_l = -8$ dB). However, for the deep fade link the modeled ellipse is considerably larger, i.e., $\lambda^+ = 0.8413$ whereas $\lambda^- = 0.0530$. Thus, we can also state that the anti-fade links indicate the person's location more accurately when increases in signal strength are measured. For example, when a link measures a $10$ dBm increase in RSS, the anti-fade link ($F_l = 8$ dB) has a 97$\%$ probability that the person is inside the modeled ellipse. In comparison, there is a 63$\%$ probability with the deep fade link ($F_l = -8$ dB).

Different models are used for the negative and positive RSS changes. The decay rates $\beta^-$ for the different fade levels when $\Delta r_{c,l} < 0$ are nearly constant thus a fixed value is used in (\ref{E:measurement_model}). On the contrary, decay rate $\beta^+$ is a function of the fade level. Thus, $\beta_{c,l}^{+}$ can be determined separately for each link and frequency channel based on the fade level of the link as follows:
\begin{equation}\label{E:pr_fit}
    \beta_{c,l}^{+} = b_{\beta}^{+} \cdot e^{F_{c,l} / k_{\beta}^{+} }.
\end{equation}
The measured and modeled decay rates are shown in Fig. \ref{F:model_fits}(b) and the parameters of the models are listed in Table \ref{T:measurement_model_parameters}.
Now, the probability of the person being located inside the modeled ellipse at time $k$ and measured RSS change $\Delta r_{c,l}$ can be estimated as:
\begin{equation}\label{E:measurement_model}
    p_{c,l}^{\delta}(k) = 1 - e^{-\beta_{c,l}^{\delta} \lvert \Delta r_{c,l}(k) \rvert}.
\end{equation}

\begin{table}
    \caption{Measurement model parameters} 
        \centering 
        \begin{tabular}{c c c} 
        \hline\hline\ 
        $\beta^{-}$  & $k_{\beta}^{+}$ & $b_{\beta}^{+}$ \\
        \hline  
        0.1172 & 13.0018 & 0.1839 \\ 
        \hline 
        \end{tabular}
        \label{T:measurement_model_parameters} 
\end{table}

\section{Methods} \label{S:model}

The image reconstruction procedure for RTI can be used as a theoretical framework for estimating the changes in the RF propagation field with the new weight and measurement models. However, minor adjustments need to be made to RTI as it was introduced in Section \ref{S:rti}. First, instead of applying the changes in RSS as given in (\ref{E:shadowing}), we apply the probability of the person being located inside the modeled ellipse which is given in (\ref{E:measurement_model}). Thus, the new measurement vector on channel $c$ when attenuation is measured is $\mathbf{y_c^-} = [p_{c,1}^{-}, \ldots, p_{c,L}^{-}]$. Correspondingly, for measured increases in RSS $\mathbf{y_c^+} = [p_{c,1}^{+}, \ldots, p_{c,L}^{+}]$. Thus, the complete measurement vector on channel $c$ becomes $\mathbf{y_c} = [\mathbf{y_c^+} \vert \mathbf{y_c^-}]$. When considering all the channels, the complete measurement vector is $\mathbf{y} = [\mathbf{y_1} \vert \ldots \vert \mathbf{y_C}]^T$, where $C$ is the number of used channels.

The spatial weighting model in (\ref{E:weight}) has to be reformulated since $\lambda$ is now unique for each link. The new model can be mathematically expressed as follows:
\begin{equation}\label{E:new_weight}
    w_{c,l,j}^{\delta} =  \begin{cases}
                       \frac{1}{n_j \cdot p^2} & \text{if } d_{lj}^{tx}+d_{lj}^{rx}<d+\lambda_{c,l}^{\delta}\\
                       0 & \text{otherwise}
                                  \end{cases},
\end{equation}
where $w_{c,l,j}^{\delta}$ is the weight of voxel $j$ for link $l$ on channel $c$ for RSS change direction $\delta$. Because the area covered by the ellipses varies, we weight less the links that cover a larger area by setting the weight to be inversely proportional to the area of the ellipse, i.e. $n_j \cdot p^2$, where $n_j$ is the number of voxels $j$ that are within the ellipse of link $l$ and $p^2$ is the area of the voxel.

Now, when all the links and channels of the RF network are considered, the changes in the RF propagation field of the monitored area can be modeled as:
\begin{equation}\label{E:new_linear_formulation}
    \mathbf{y} =  \mathbf{W} \mathbf{x} + \mathbf{n} ,
\end{equation}
where $\mathbf{y}$ and $\mathbf{n}$ are the measurement and noise vectors, correspondingly, both of size $2LC \times 1$. As in (\ref{E:linear_model}), $\mathbf{x}$ is the image to be estimated, and $\mathbf{W}$ is the new multi-scale weight matrix, of size $2LC \times N$. The regularized least-squares approach in (\ref{E:regularization}) can be used also with the new models by substituting the weight matrix $\mathbf{A}$ with the new weight matrix $\mathbf{W}$.

\subsection{Localization and Tracking} \label{S:localization_and_tracking}

From the estimated image (\ref{E:linear_transformation}), the position of a person can be determined by finding the voxel of the image that has the maximum value:
\begin{equation}\label{E:location_est}
    j =  \underset{N}{\arg \max } ~ \mathbf{\hat{x}}.
\end{equation}
The position estimate is therefore $\hat{z} = z_{j}$, where $z_{j}$ represents the coordinate of the center of voxel $j$. For tracking the movements of the person, we apply a Kalman filter \cite{barshalom2001} designed to estimate the position, velocity, and acceleration of the tracked person as in \cite{kaltiokallio2011}. Note that the localization proposed in (\ref{E:location_est}) is only capable of locating one person. Multi-target localization and tracking is a more challenging issue in DFL, and it is outside the scope of this paper. Thus, we do not propose coordinate estimators for the multi-target case. Readers are referred to \cite{wilson11fade,zhang2009dynamic,thouin2011,nannuru2011} for RSS-based multi-target tracking.

In Section \ref{S:results} we evaluate the accuracy of the system. For that purpose the localization error at time $k$ is defined as the distance between the estimated and true position:
\begin{equation}\label{E:localization_error}
    e_{k} =  \lVert \hat{z}_k - z_{H,k}\rVert,
\end{equation}
where $z_{H,k}$ is the true human location.

\subsection{Performance Benchmarking} \label{S:performance_benchmarking}

The system presented in this paper, which we denote as multi-scale RTI (msRTI), is compared to two other systems: an RTI system that ranks the channels based on the fade level \cite{kaltiokallio2012a,kaltiokallio2012b} and a RTI (see Section \ref{S:rti}) system that exploits channel diversity. The fade level RTI (flRTI) system ranks the different channels based on their fade level and uses only the measurements of the $m$ most anti-fade channels. The RSS measurements collected on the selected channels are averaged to obtain the measurement vector $\mathbf{y}$. The channel diversity RTI (cdRTI) system considers each channel as an independent link, i.e., as if at each node's location there were multiple sensors communicating on different channels. For the two systems used for benchmarking, the image reconstruction parameters that maximize the localization accuracy are derived through simulations. For the msRTI system proposed in this paper, the image reconstruction parameters, shown in Table \ref{T:ImageParameters}, are the same for all the tests.

\begin{table}
    \caption{Image reconstruction parameters used in the experiments} 
        \centering 
        \begin{tabular}{c c c} 
        \hline\hline\ 
        Parameter & Value & Description \\
        \hline  
        $p$ & 0.1524 & Voxel width [m] \\ 
        $\sigma_{x}$ & 0.0316 & Voxels standard deviation [dB]  \\ 
        $\sigma_{N}$ & 1.4142 & Noise standard deviation [dB]  \\ 
        $\delta_{c}$ & 4 & Decorrelation distance [m] \\ 
        \hline 
        \end{tabular}
        \label{T:ImageParameters} 
\end{table}

\subsection{Experiment description} \label{experiment_description}

In the beginning of the tests (see Section \ref{S:measurements}), a one minute calibration period is performed. During this period, the mean RSS and fade level of every link and channel are calculated. Online training and calibration of the network are not within the scope of this paper. However, both are research topics that will be addressed in future work. For now, readers are referred to \cite{wilson10see,zheng2012,kaltiokallio2012b} for systems that do not require calibration \cite{wilson10see}, have online training \cite{zheng2012}, or are capable of adapting to the changing environment \cite{kaltiokallio2012b}.

Experiment 1 is solely a tracking experiment. In the test a person enters the monitored area, after the initial calibration, and walks along a predefined rectangular path, covering $14$ laps before leaving the area. In experiment 2, both localization and tracking accuracy are evaluated. In the localization experiment, a person is standing still at one of 15 known positions for a predetermined period of time so to compute multiple position estimates at each location. The 15 positions are chosen so as to cover all the areas of the apartment. During the experiment, each position is visited multiple times. In the tracking experiment, a person is moving at constant speed along the paths connecting the same 15 locations used in the localization experiment.

Since the models presented in this paper are derived from the data collected in the environments of experiments 1 and 2, it is important to validate their generality in an environment differing from the ones used to build them. Hence, we conduct an additional test in a more challenging through-wall scenario (70m$^2$). In the experiment, 30 nodes are deployed around a lounge room, outside the walls surrounding the monitored area. The sensors are set on podiums as in experiment 1. Five channels (11, 15, 18, 21, and 26) are used for communication. During the experiment, the person is standing still at one of ten predefined positions, located along a rectangular perimeter so to evenly cover the entire monitored area. Besides evaluating the localization accuracy of the system in a through-wall scenario, we also focus on analyzing the effect that the number of frequency channels used for communication has on the accuracy.

\section{Experimental Results} \label{S:results}

\subsection{Experiment 1, open environment}
\begin{figure*}
\mbox{
      \subfigure[\quad]{\includegraphics[width=2.2in]{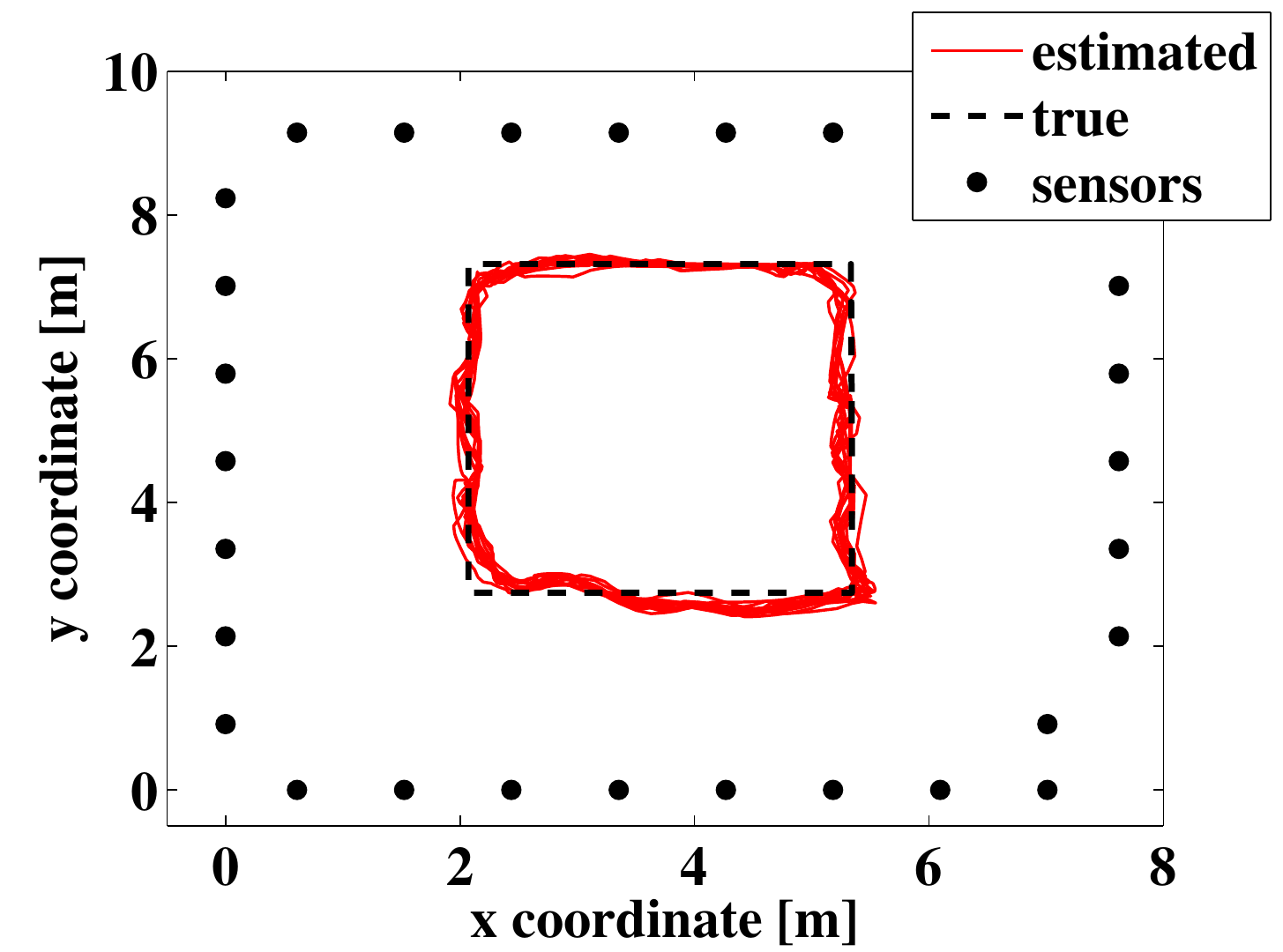}} \quad
      \subfigure[\quad]{\includegraphics[width=2.2in]{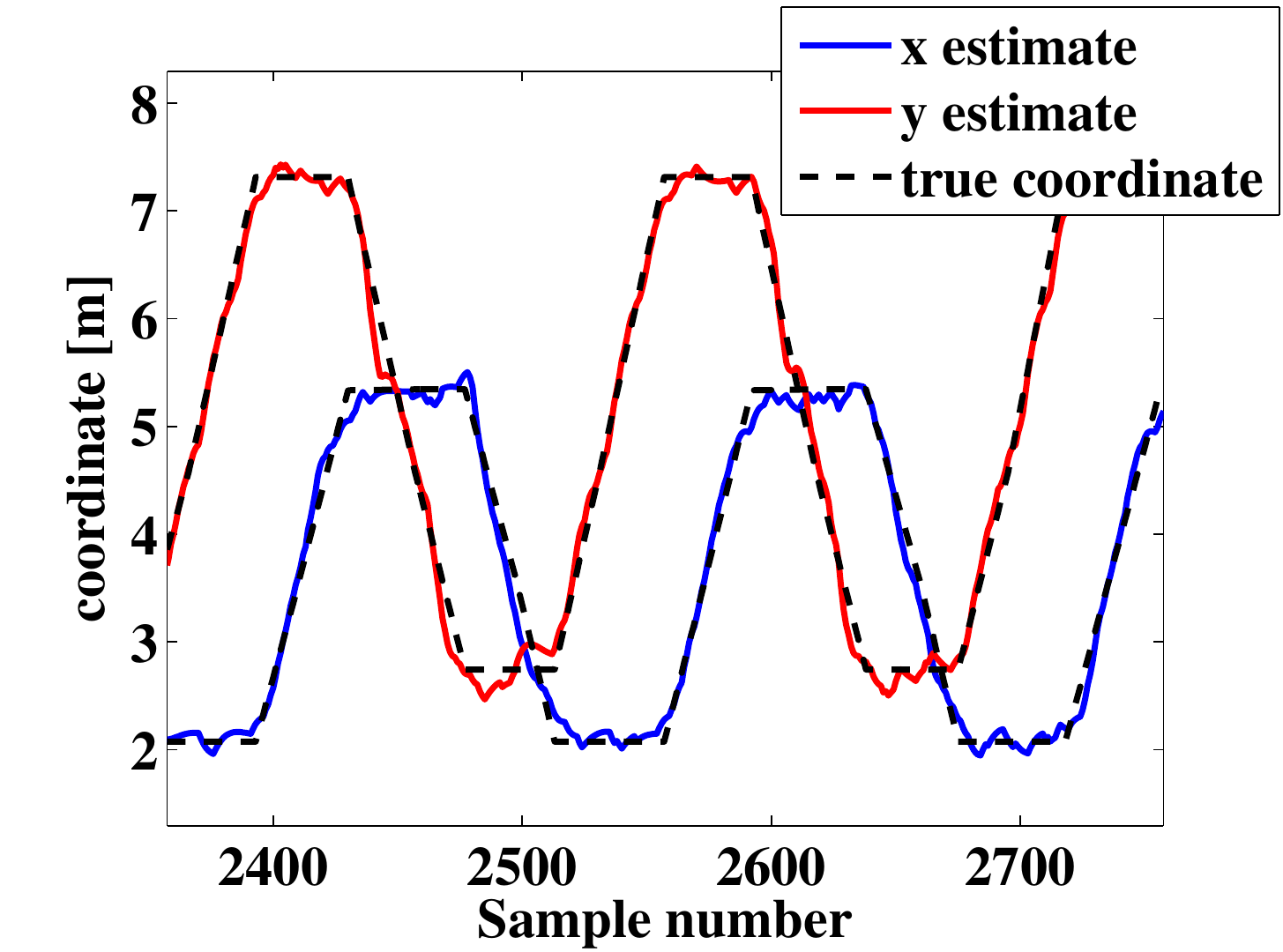}} \quad
      \subfigure[\quad]{\includegraphics[width=2.2in]{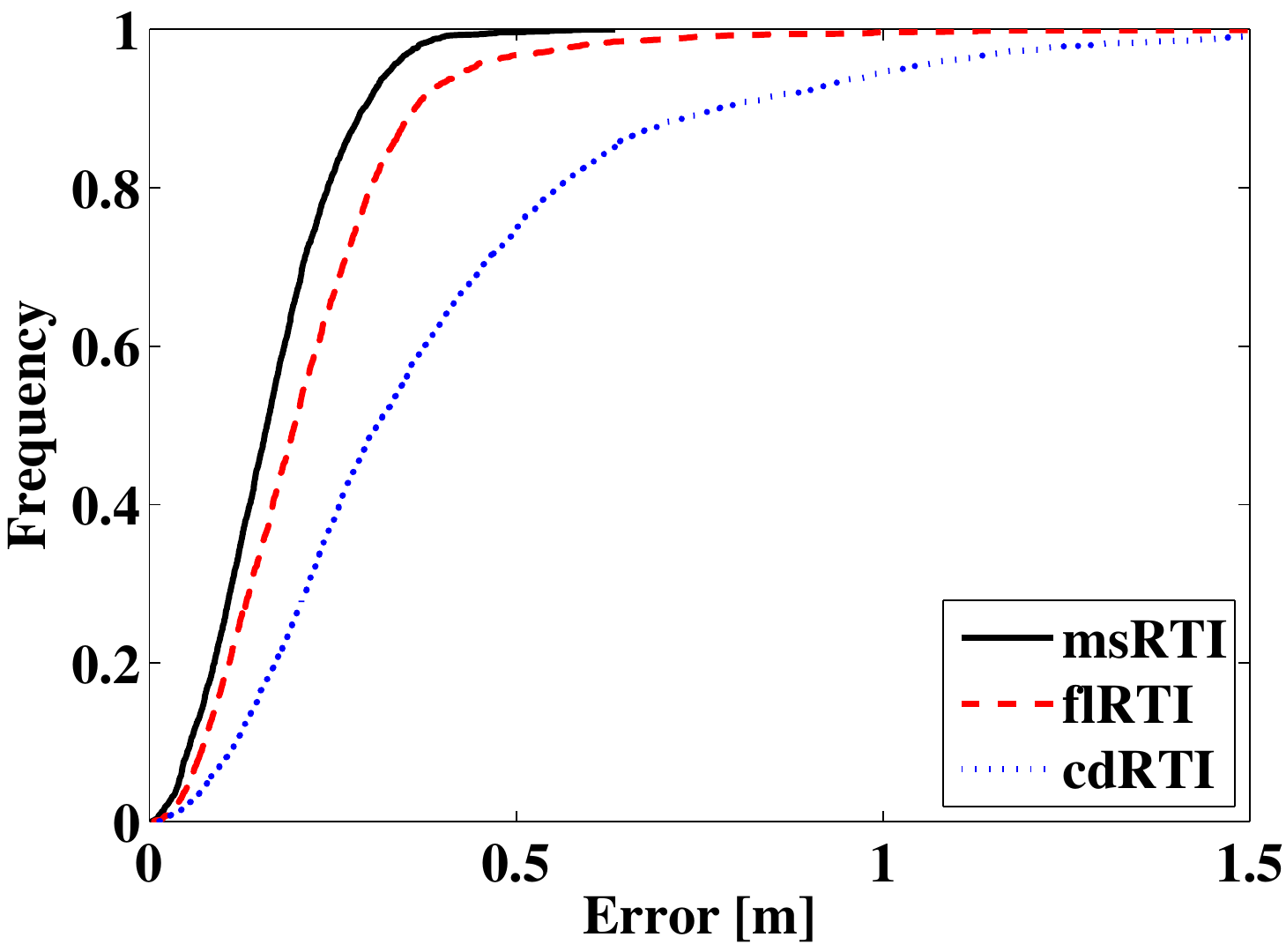}} \quad
      }
    \caption{In (a), the estimated and true trajectories of the person in the open environment test and in (b), the coordinate estimates of the person over two consecutive laps. In (c), CDFs of the tracking accuracy with the three different methods in the open environment experiment.}
    \label{F:experiment1}
\end{figure*}

\begin{figure*}
    \mbox{
      \subfigure[\quad]{\includegraphics[width=2.2in]{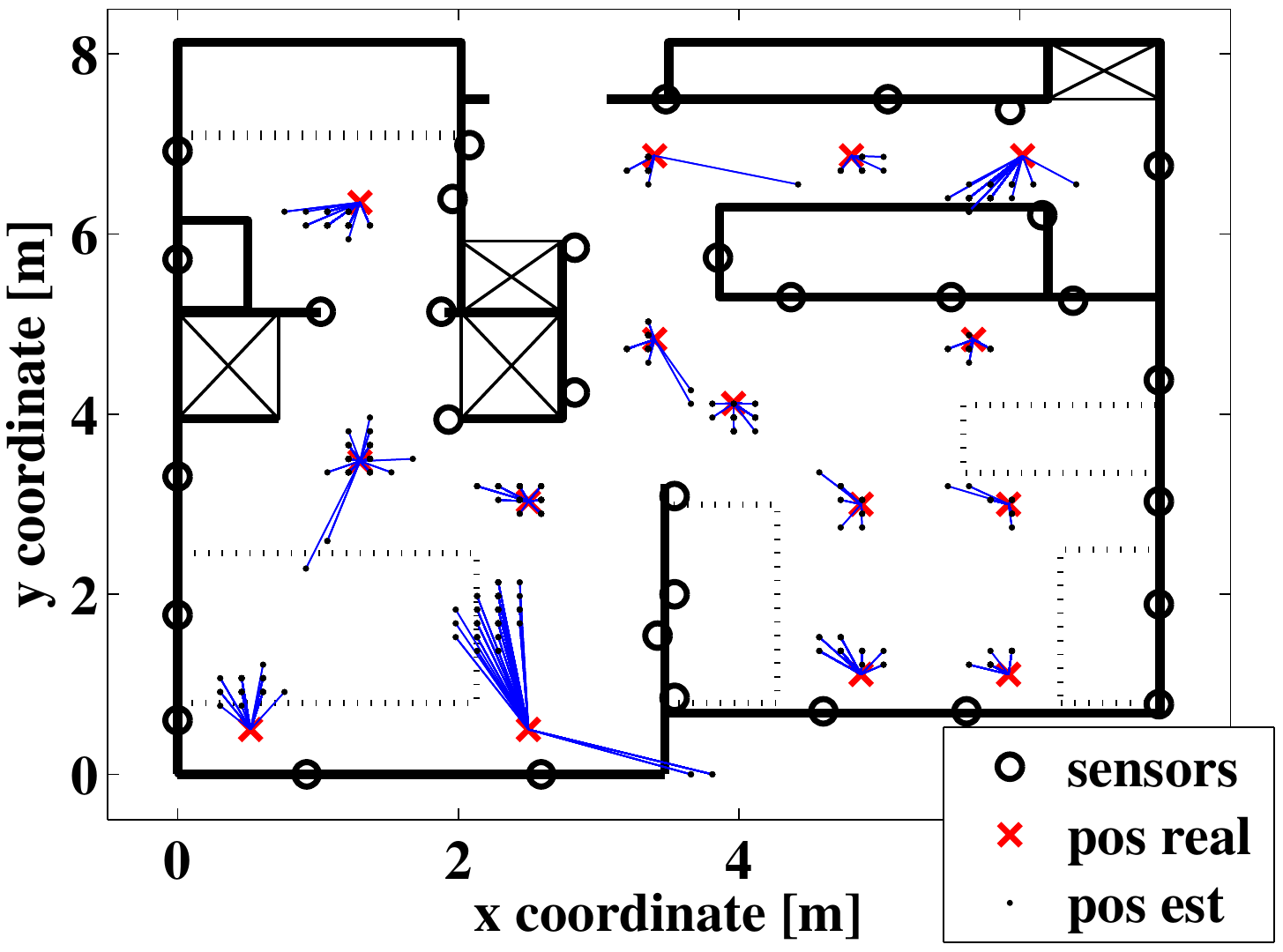}} \quad
      \subfigure[\quad]{\includegraphics[width=2.2in]{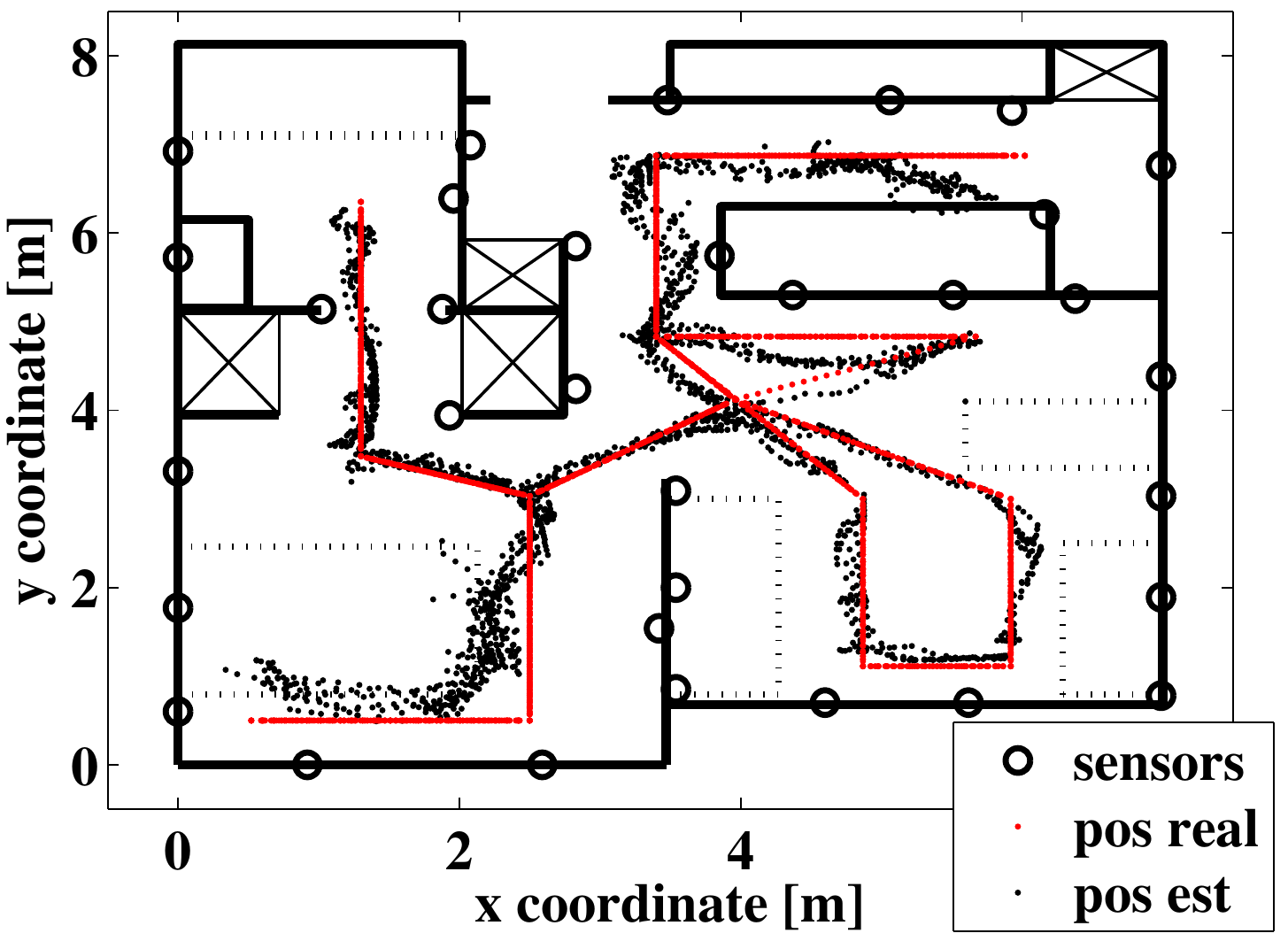}} \quad
      \subfigure[\quad]{\includegraphics[width=2.2in]{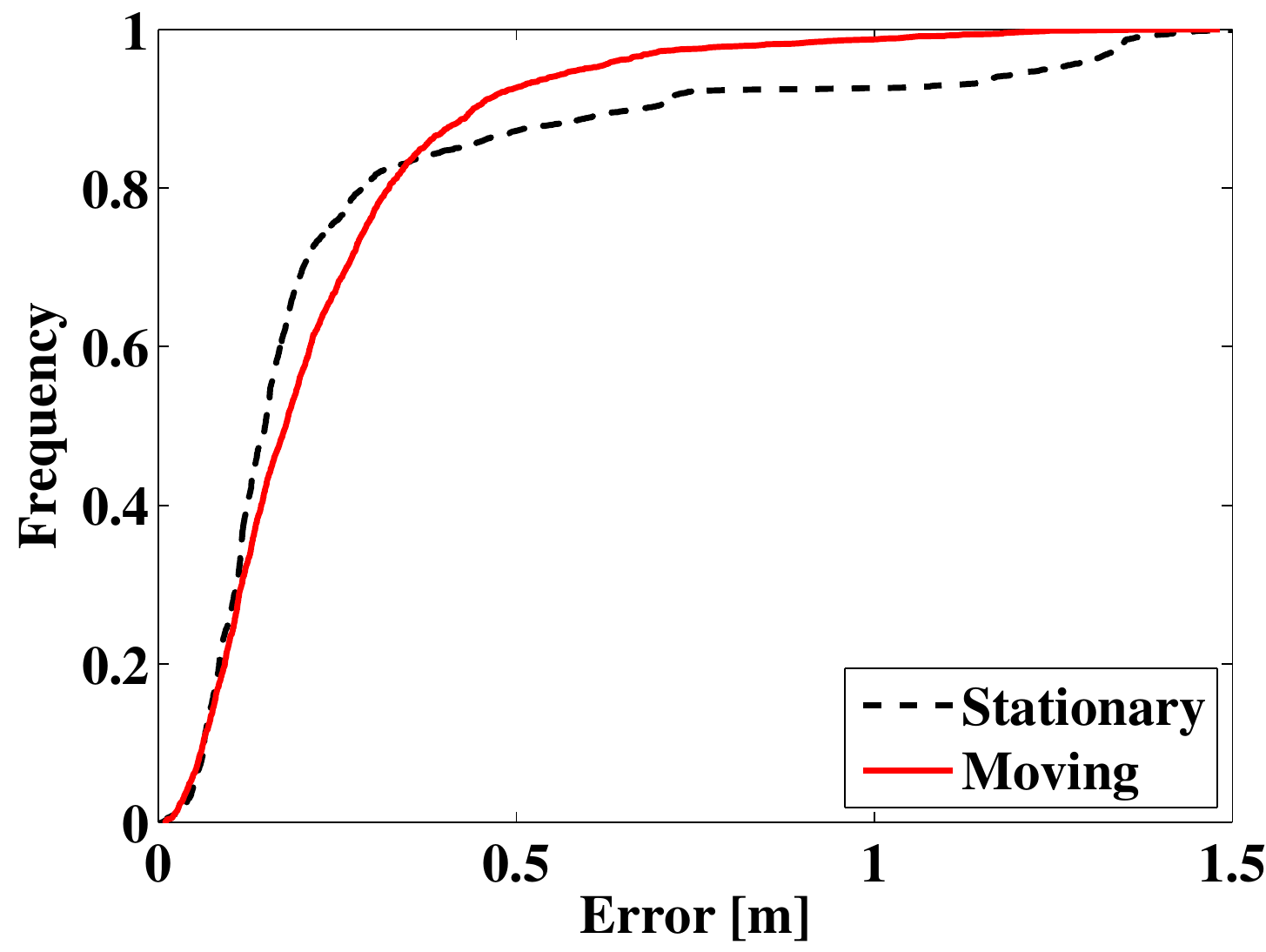}} \quad
      }
    \caption{In (a), the estimated and true locations of a stationary person. In (b), the estimated and true trajectories of a moving person. In (c), CDFs of locating a stationary and moving person with msRTI in the apartment experiment.}
    \label{F:experiment2}
\end{figure*}

The true and estimated trajectory of the person during the test are shown in Fig. \ref{F:experiment1}(a). The coordinate's estimates of two consecutive laps of the test are shown in Fig. \ref{F:experiment1}(b). The msRTI system is capable of tracking the person with an average error of $\bar{e}=0.17$ m, whereas the average error for flRTI and cdRTI systems are $\bar{e}=0.22$ m and $\bar{e}=0.39$ m respectively. In \cite{kaltiokallio2012a}, we demonstrate that an RTI system exploiting channel diversity always outperforms an RTI system relying on single channel communication and that the performance can be increased even further when the channels are ranked based on their fade level. However, the drawback of the method used in the flRTI system is that the RSS measurements collected on the deep fade channels are actually not considered at all. In fact, when the traditional spatial weight model is applied, not considering these noisy data improves the localization accuracy. Instead, when the new methods are applied, the RSS measured on all the channels can be effectively used to further improve the localization accuracy, as demonstrated by the results.

The cumulative distribution functions (CDFs) of the localization errors for the three different systems are shown in Fig. \ref{F:experiment1}(c). In this experiment, the median error is $0.16$ m with msRTI, $0.20$ m with flRTI and $0.30$ m with cdRTI. One major advantage of using msRTI is that all the position estimates are accurate $-$ the most inaccurate being 0.63 m. With the other two systems, some position estimates are more than a meter off the true position of the person. The highest errors for the two methods are 2.03 m with flRTI and 2.12 m with cdRTI.

\subsection{Experiment 2, apartment}

The true and estimated positions of the localization experiment are shown in Fig. \ref{F:experiment2}(a). Respectively, the true and estimated trajectories are shown in Fig. \ref{F:experiment2}(b). The average localization accuracy of the experiment is $\bar{e}=0.26$ m and the average tracking accuracy of the experiment is $\bar{e}=0.23$ m with msRTI. The CDFs of the two experiments are shown in Fig. \ref{F:experiment2}(c). The median error of the localization and tracking experiments are 0.15 m and 0.18 m, respectively. Correspondingly, the 95th-percentile of the two errors are 1.24 m (localization) and 0.59 m (tracking). The location found in the bedroom degrades considerably the accuracy of both experiments. However, since in the localization experiment the time spent in that particular location is longer, the localization results are affected more than the tracking ones.

The combined error of the localization and tracking experiments is 0.25 m with msRTI, whereas it is 0.26 m with flRTI and 0.36 m with cdRTI. The difference in accuracy among the systems is not as remarkable as in the open environment. In the apartment, the sensors are deployed in all the rooms, and there exist many sensors that have LoS communication with each other. In addition, the distance separating some of the sensors is small. Due to this, the links measure on average attenuation as the link line is obstructed. Therefore, an attenuation-based system that exploits a fixed $\lambda$ model is capable of performing adequately well in the apartment deployment.

\subsection{Experiment 3, through-wall}

\begin{figure*}
    \mbox{
      \subfigure[\quad]{\includegraphics[width=\columnwidth]{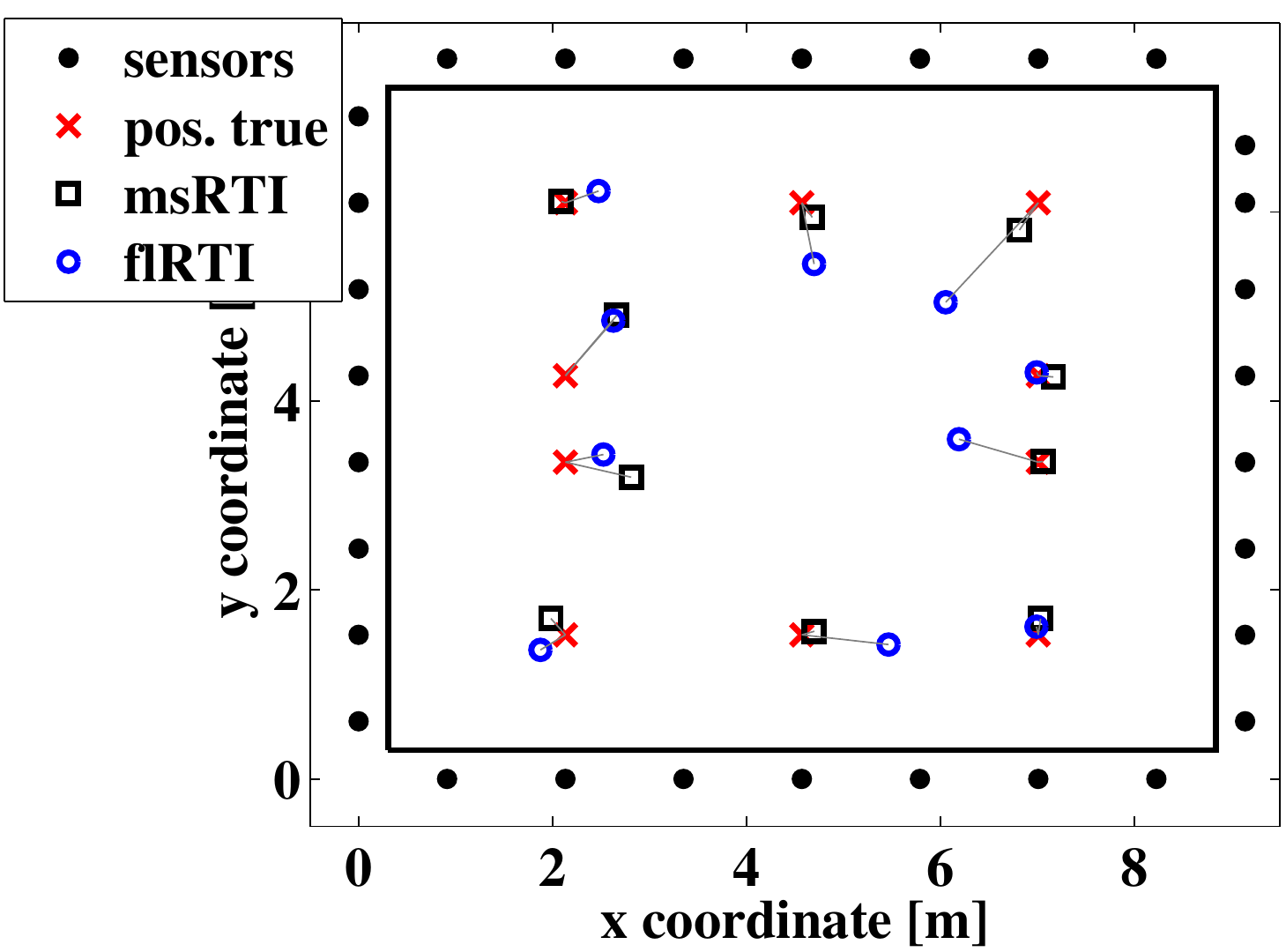}} \quad
      \subfigure[\quad]{\includegraphics[width=\columnwidth]{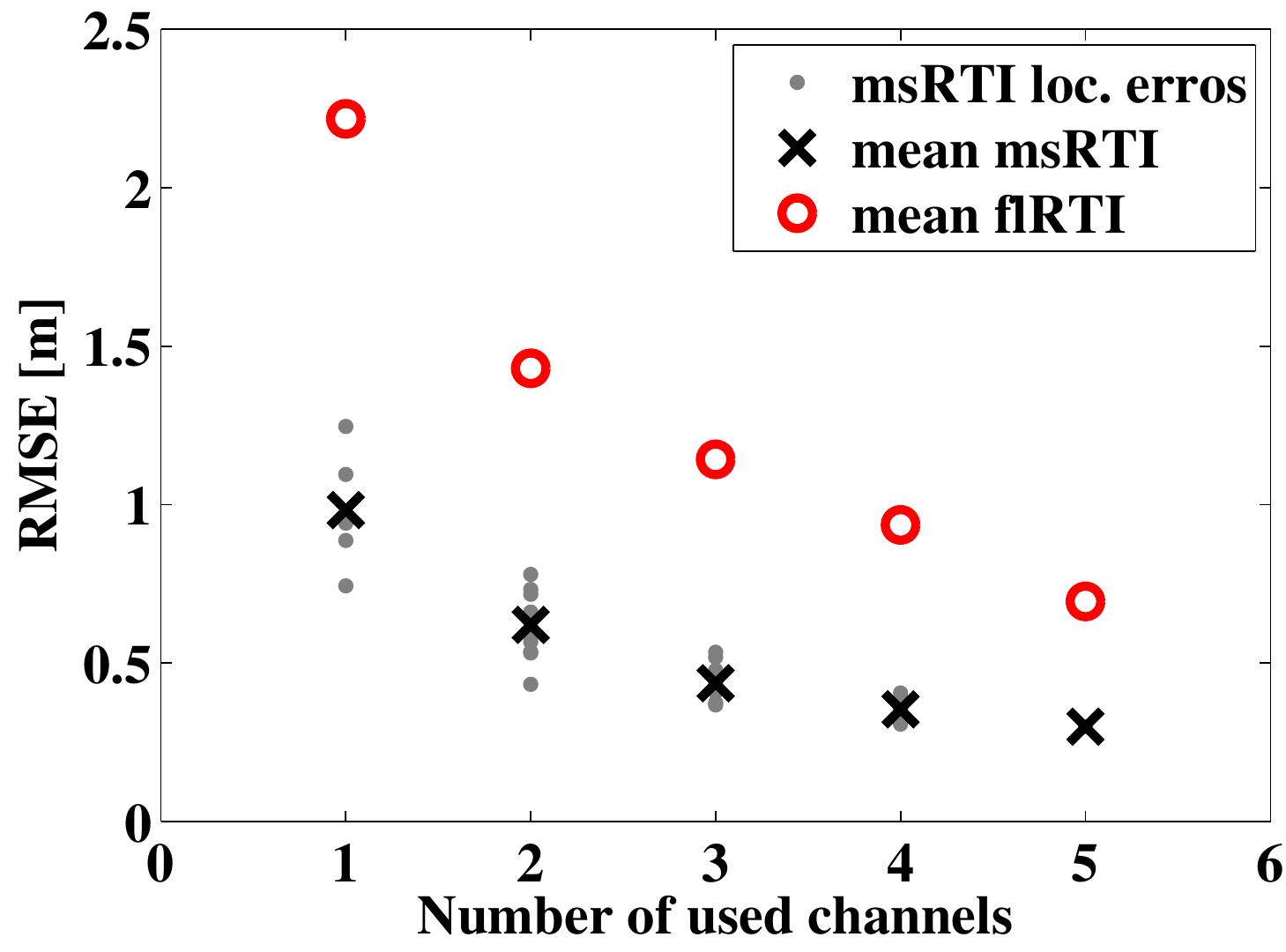}} \quad
      }
    \caption{The estimated and true positions of the person with msRTI and flRTI is shown in (a). The effect of channel number on the localization accuracy of the systems presented in (b).}
    \label{F:experiment3}
\end{figure*}

\begin{figure*}
   \mbox{
      \subfigure[\quad]{\includegraphics[width=1.6in]{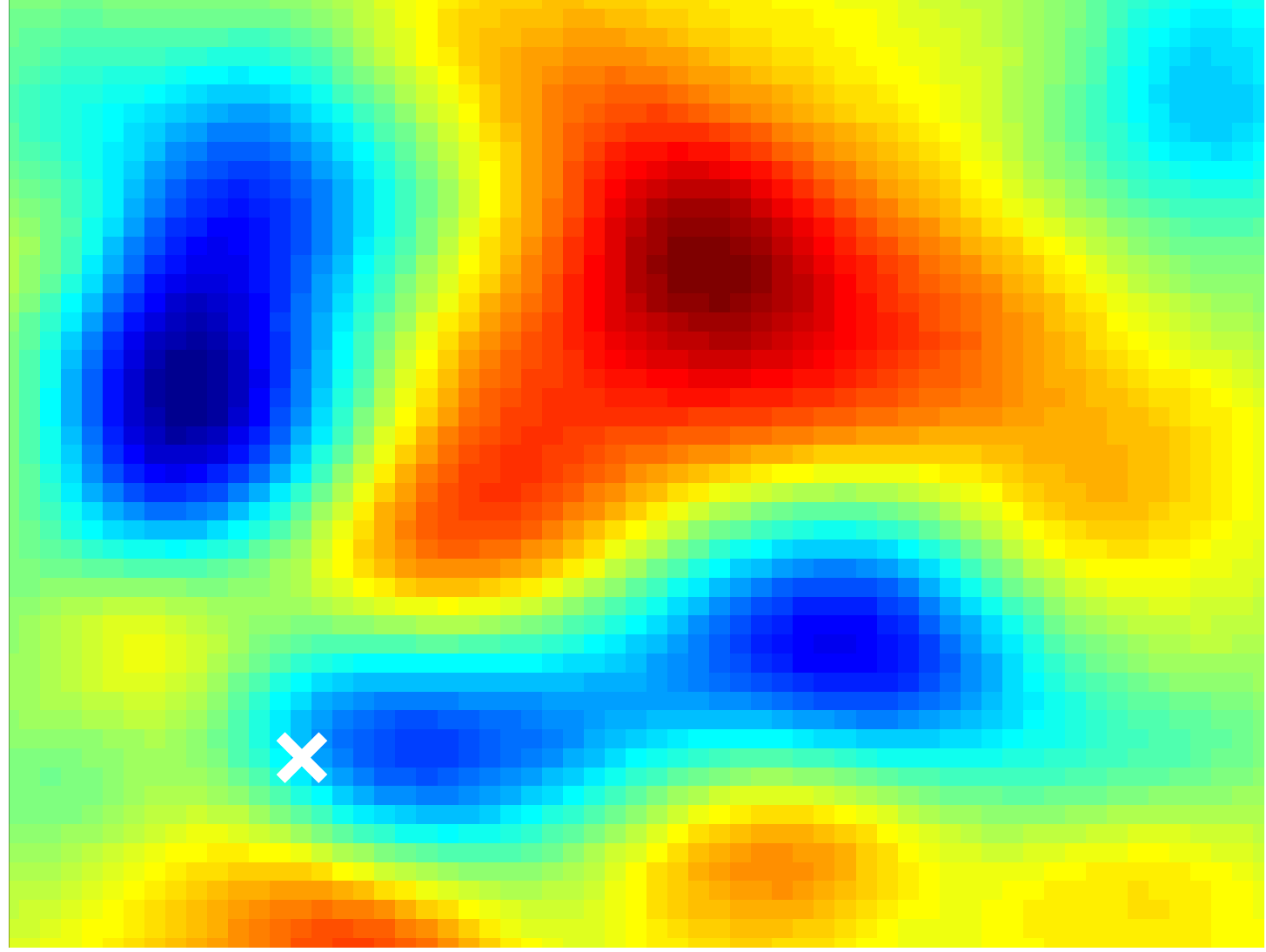}} \quad
      \subfigure[\quad]{\includegraphics[width=1.6in]{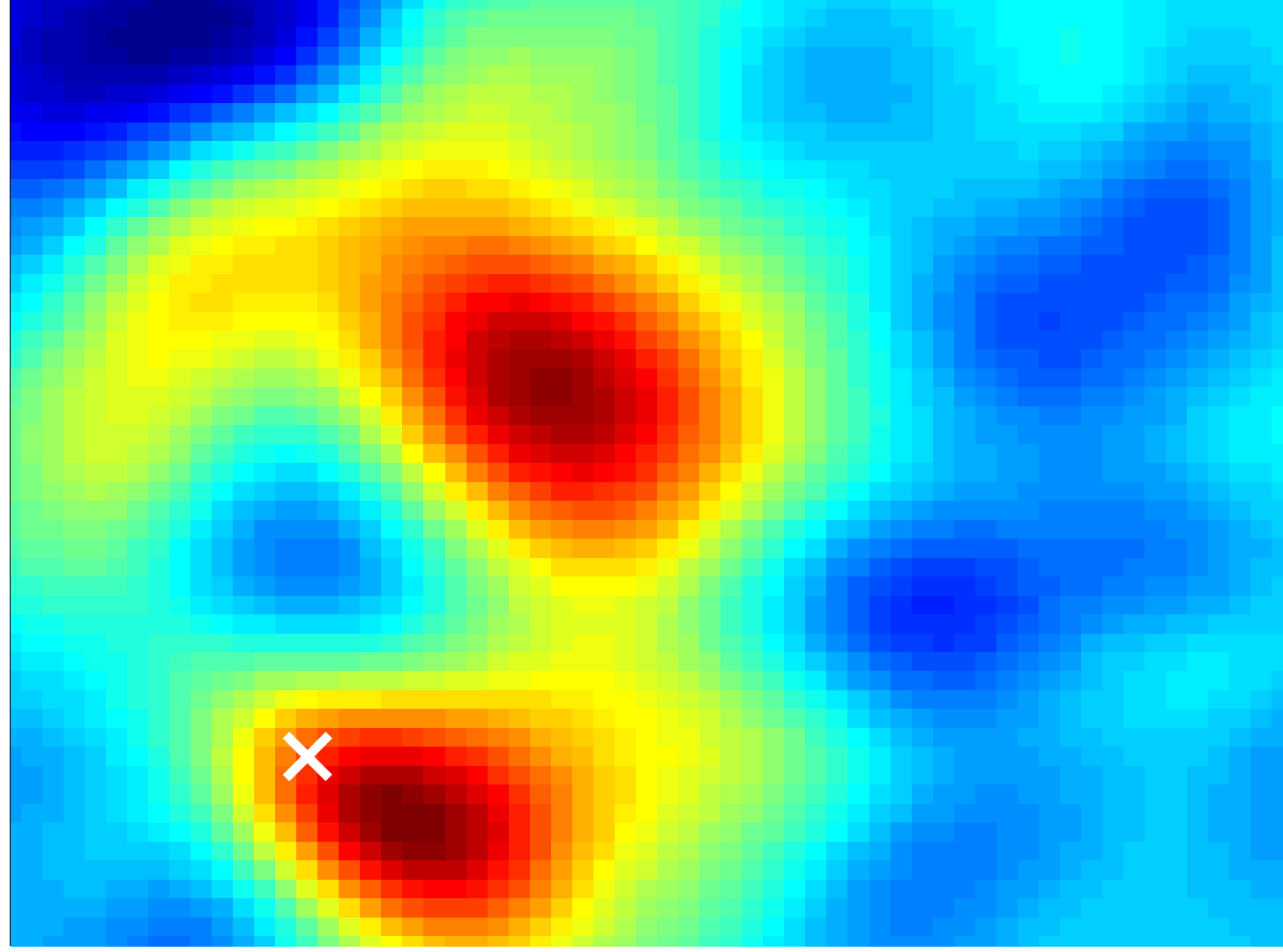}} \quad
      \subfigure[\quad]{\includegraphics[width=1.6in]{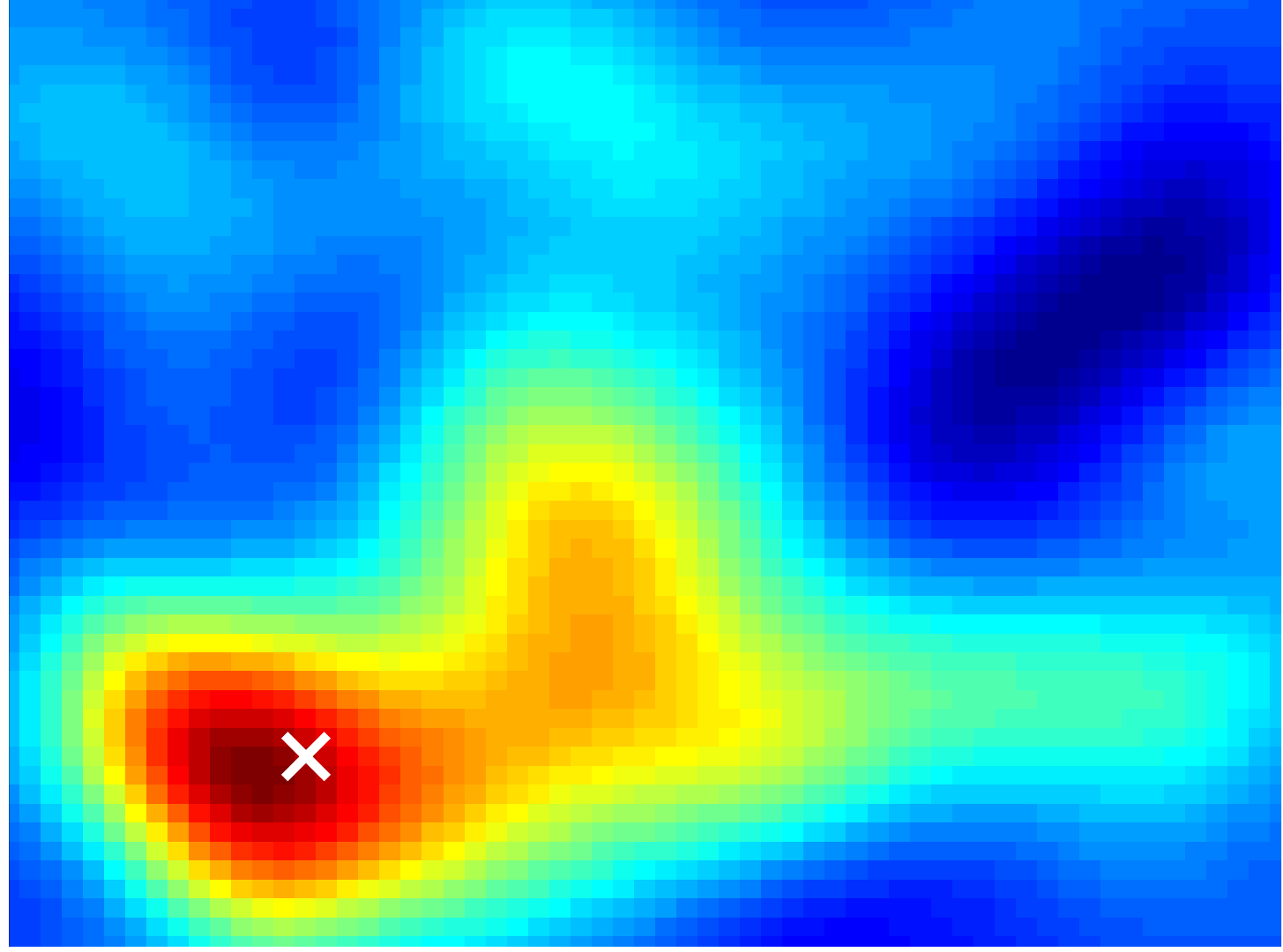}} \quad
      \subfigure[\quad]{\includegraphics[width=1.6in]{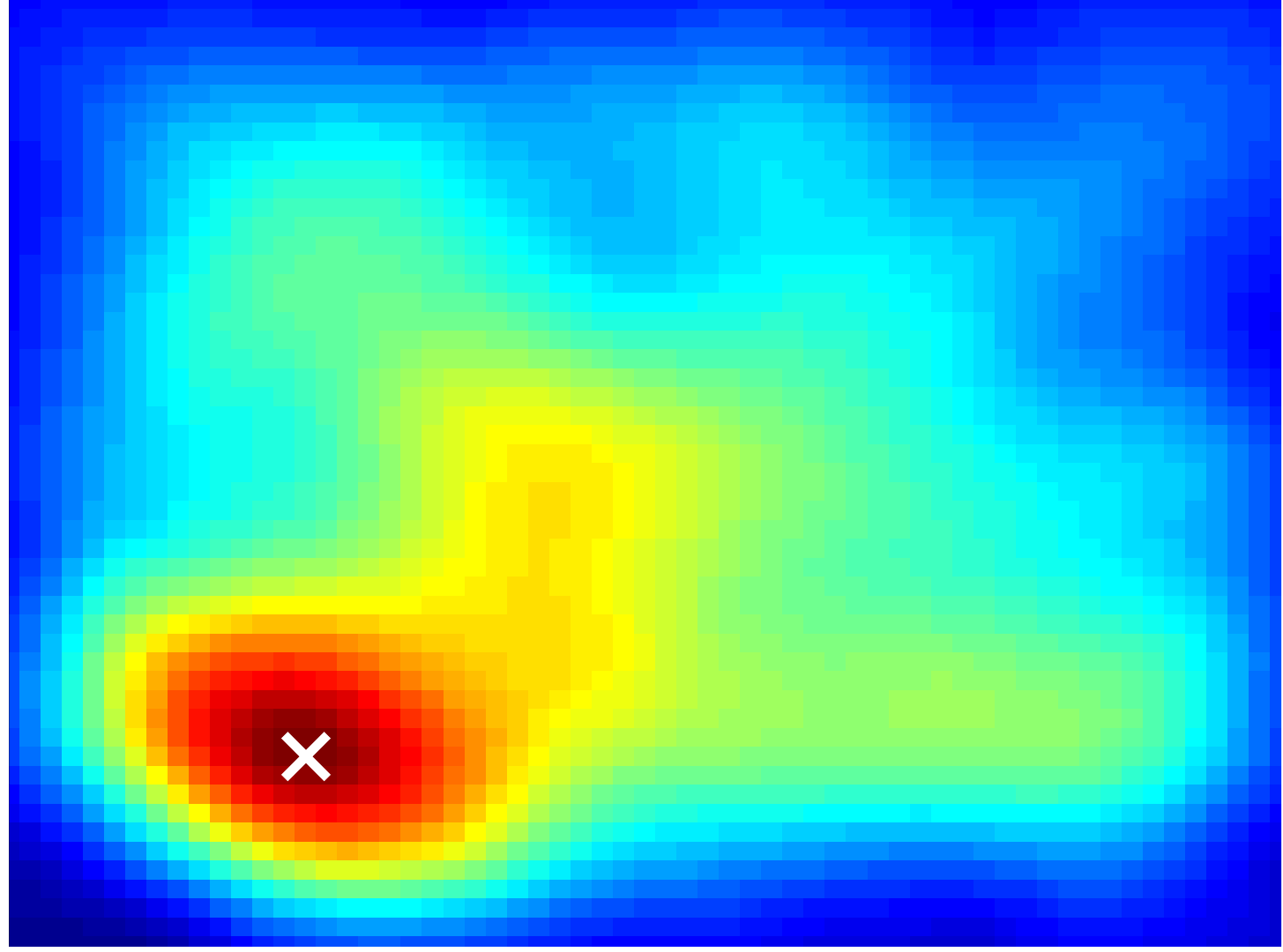}} \quad
      }

    \caption{The estimated change in the attenuation field with four different methods. In (a), using RTI on a single channel. In (b), using cdRTI (each channel is considered as a unique link). In (c), the channels are ranked based on their fade level and the three most anti-fade channels are used (flRTI). In (d), with the new models presented in this paper (msRTI). The true location of the person is represented by the white cross.}
    \label{F:propagationFields}
\end{figure*}

The true and estimated positions of the experiment are shown in Fig. \ref{F:experiment3}(a). The figure shows for comparison the results obtained with flRTI. The average localization error is $\bar{e}=0.30$ m with msRTI, whereas with flRTI the average localization accuracy is $\bar{e}=0.70$ m, and $\bar{e}=1.57$ m with cdRTI. For clarity, the results of cdRTI are omitted from Fig \ref{F:experiment3}(a). The improved performance of the new models is due to the fact that in the through-wall scenario the number of links measuring attenuation when the link line is obstructed is considerably smaller than in the other two environments. Furthermore, in this type of environment it is common that human-induced changes in the RSS are observed also elsewhere than on the link line. This further motivates the use of the multi-scale spatial weight model in cluttered environments.

As demonstrated in \cite{kaltiokallio2012a}, channel diversity enhances the accuracy of DFL considerably, since it increases the probability that at least one (or more) of the channels will be in anti-fade, thus favoring the localization attempt. Figure \ref{F:experiment3}(b) shows the average localization errors over the ten predetermined locations obtained with msRTI for all the possible combinations of the five channels used in the through-wall experiment. The average localization error obtained with msRTI is smaller than the ones obtained with flRTI. The msRTI system achieves a better average accuracy by using only two channels than the flRTI by using all five channels.

The advantage of exploiting channel diversity and the new models is shown in Fig. \ref{F:propagationFields}. The figure shows the estimated change in the propagation field with four different methods. In (a), the change in the attenuation field is estimated by using a single channel for communication (RTI). With this method, the localization error is $5$ m. In (b), channel diversity is exploited (cdRTI). Even though the position estimate obtained with this method is accurate, the image is very noisy, containing two separate "blobs" which could be interpreted as two different individuals. In (c), the change in the attenuation field is estimated by ranking the channels based on their fade level and using only the three most anti-fade channels (flRTI). The system locates the person with a 0.34 cm error. Not only is the image more accurate, it also contains much less noise then in (a) and (b). In (d), the new models are applied (msRTI). The true and estimated positions overlap, and the image is not corrupted by noise.


\section{Conclusion}
In this paper, we present new methods to enhance the accuracy of RSS-based DFL. The improvements concern four aspects: exploiting channel diversity, deriving a more accurate spatial model for the human-induced RSS changes, proposing a measurement model that determines the probability of the person being inside the modeled area, and taking into consideration the direction of the RSS change. In the paper, we experimentally show that a person located on the link line not only causes attenuation of the signal but can also cause the signal strength to increase. In addition, we show that a person can influence the link measurements even far away from the link line. Both results are found to depend on the fade level of the link$-$ a measure indicating if the link is experiencing destructive or constructive multipath interference. The fade level is then used to derive more accurate measurement and spatial models for the human-induced changes in the RSS.

The performance of the new models proposed in this paper are validated in three different indoor environments (open, cluttered, through-wall). The results demonstrate that the new method outperforms the current state-of-the-art methods in \cite{wilson09a,kaltiokallio2012a}. Moreover, the improvement in localization accuracy with the new system is more consistent the more the environment is challenging for RSS-based DFL. The results indicate that the system here presented is the most accurate RSS-based DFL system reported to date, capable of achieving 0.30 m localization accuracy even in through-wall scenarios.

In future work, we will investigate spatial propagation models that can adaptively learn the geometrical propagation patterns of the multipath components and their associated amplitudes. Adapting the models to the environment can provide the means to further improve the accuracy of DFL. So far, only the spatial correlations of the link measurements have been considered when estimating the changes in the propagation field. However, the link measurements are also time correlated. Future research will explore how to exploit the time correlation of the measurements in the context of DFL.


\section*{Acknowledgments}
This work is funded by the Finnish Funding Agency for Technology and Innovation (TEKES). This material is also based upon work supported by the U.S National Science Foundation under grants 0748206 and 1035565. The authors wish to thank Brad Mager for the help in setting up the experiments.
\addcontentsline{toc}{section}{Acknowledgment}


\begin{biography}[{\includegraphics[width=1.0in,height=1.25in]{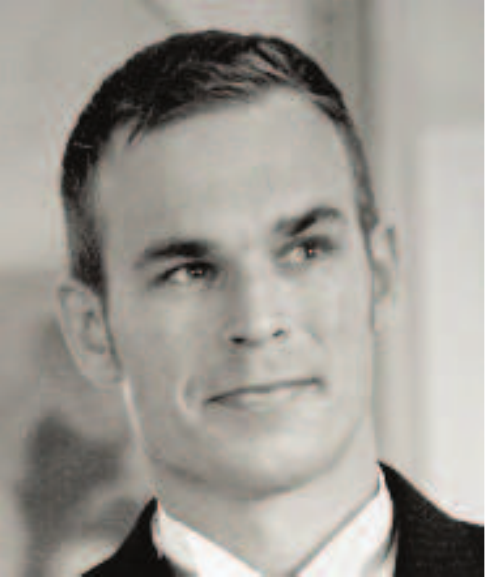}}]{Ossi Kaltiokallio}
received the B.Sc and M.Sc degrees in electrical engineering from Aalto University, School of Electrical Engineering, Helsinki, Finland, both in 2011. He is currently a Ph.D. student with the Department of Automation and Systems Technology, Aalto University School of Electrical Engineering. He is a member of the Wireless Sensor Systems Group at Aalto University. His current research interests include RSS based localization; signal processing, and design and implementation of embedded wireless systems.
\end{biography}
\begin{biography}[{\includegraphics[width=1.0in,height=1.25in]{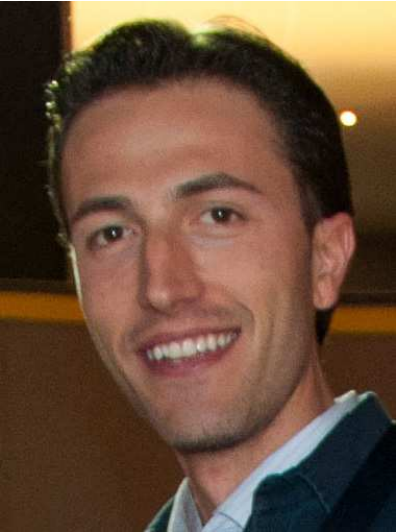}}]{Maurizio Bocca}
received the B.Sc. (2003) and M.Sc. (2006) degrees in Computer Science Engineering from the Politecnico di Milano (Milan, Italy), and the Ph.D. (2011) in Electrical Engineering from Aalto University (Helsinki, Finland). In 2012, he has joined as a post doc the Sensing and Processing Across Networks (SPAN) Lab at the University of Utah (Salt Lake City, Utah, USA), where he is conducting research in the area of RF sensor networks for device-free localization, context awareness and elder care. His research interests include distributed and adaptive algorithms for wireless sensor networks and smart protocols for large-scale deployments of sensor networks in real-world scenarios.
\end{biography}
\begin{biography}[{\includegraphics[width=1.0in,height=1.25in]{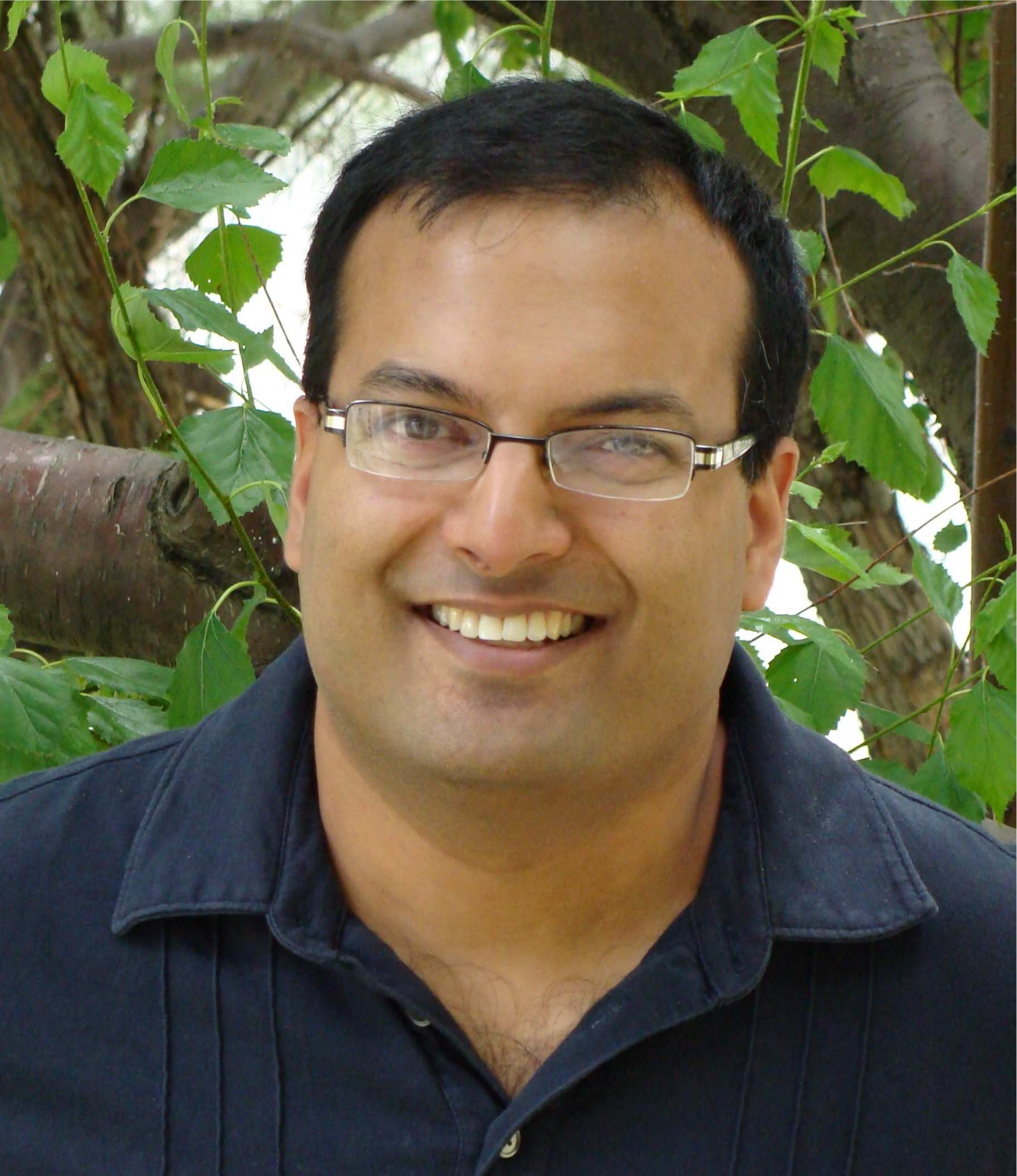}}]{Neal Patwari}
Neal Patwari received the B.S. (1997) and M.S. (1999) degrees from Virginia Tech, and the Ph.D. from the University of Michigan, Ann Arbor (2005), all in Electrical Engineering. He was a research engineer in Motorola Labs, Florida, between 1999 and 2001.  Since 2006, he has been at the University of Utah, where he is an Associate Professor in the Department of Electrical and Computer Engineering, with an adjunct appointment in the School of Computing.  He directs the Sensing and Processing Across Networks (SPAN) Lab, which performs research at the intersection of statistical signal processing and wireless networking. Neal is the Director of Research at Xandem, a Salt Lake City-based technology company.  His research interests are in radio channel signal processing, in which radio channel measurements are used to benefit security, networking, and localization applications.  He received the NSF CAREER Award in 2008, the 2009 IEEE Signal Processing Society Best Magazine Paper Award, and the 2011 University of Utah Early Career Teaching Award. He is an associate editor of the IEEE Transactions on Mobile Computing.
\end{biography}

\end{document}